\documentclass[journal=jacsat,manuscript=article]{achemso}
\usepackage{graphicx}
\usepackage{amsmath}
\usepackage{subcaption}
\usepackage{float}
\usepackage{textcomp}
\usepackage{textgreek}
\usepackage{xcolor}
\usepackage{tablefootnote}
\usepackage{url}
\usepackage{epstopdf}
\usepackage{xr}
\usepackage{multirow}
\usepackage{nicefrac}
\usepackage{lscape}

\DeclareUnicodeCharacter{2009}{FIXME}

\makeatletter
\newcommand*{\addFileDependency}[1]{
  \typeout{(#1)}
  \@addtofilelist{#1}
  \IfFileExists{#1}{}{\typeout{No file #1.}}
}
\makeatother

\newcommand*{\myexternaldocument}[1]{%
	\externaldocument{#1}%
	\addFileDependency{#1.tex}%
	\addFileDependency{#1.aux}%
}

\myexternaldocument{si}

\title
  {How does the preparation strategy influence the gold cap of thermophoretic Janus swimmers and their propulsion?}

\author{Franziska Braun}
\affiliation{Soft Matter at Interfaces, Institute for Condensed Matter Physics, Technische Universität Darmstadt, Darmstadt, Germany}
\author{Michael Florian Peter Wagner}
\affiliation{Materials Research, GSI Helmholtzzentrum für Schwerionenforschung, Darmstadt, Germany}
\author{Maria Eugenia Toimil-Molares}
\affiliation{Materials Research, GSI Helmholtzzentrum für Schwerionenforschung, Darmstadt, Germany}
\author{Regine von Klitzing}
\affiliation{Soft Matter at Interfaces, Institute for Condensed Matter Physics, Technische Universität Darmstadt, Darmstadt, Germany}
\email{klitzing@smi.tu-darmstadt.de}

\affiliation[Soft Matter at Interfaces, Institute for Condensed Matter Physics, Technical University of Darmstadt, Darmstadt, Germany]
{Soft Matter at Interfaces, Institute for Condensed Matter Physics, Technische Universität Darmstadt, Darmstadt, Germany}

\begin{document}
\maketitle
\begin{abstract}
\noindent The motion of partly gold (Au)-coated Janus particles under laser irradiation is caused by self-thermophoresis. Despite numerous studies addressing this topic, the impact of the preparation method and the degree of coverage of the particle with Au on the resulting thermophoretic velocity has not yet been fully understood. A detailed understanding of the most important tuning parameters during the preparation process is crucial to design Janus particles that are optimized in Au coverage to receive a high thermophoretic velocity. In this study, we explore the influence of the fabrication process, which changes the Au cap size, on the resulting self-propulsion behavior of partly Au-coated polystyrene particles (Au-PS). Additionally, also the impact of an underlying adhesion chromium layer is investigated. In addition to the most commonly used qualitative SEM and EDX measurements, we propose a novel technique utilizing AFM studies to quantify the cap size. This non-invasive technique can be used to determine both the size and the maximum thickness of the Au cap. The Au cap size was systematically varied in a range between about 36\,\% and 74\,\% by different preparation strategies. Nevertheless, we showed that the differing Au cap size of the Janus particles in this range has no effect on the thermophoretic velocity. This surprising result is discussed in the paper.
\end{abstract}

\section{1. Introduction}
Janus particles are named after the double-faced Roman god "Janus" since they combine different chemical or physical properties at their opposite sides\cite{deGennes1992}. During the last two decades, Janus particles have gained increasing attention due to a variety of possible applications, e.g., as emulsions stabilizers\cite{Walther2008,Aveyard2012}, in biological sciences\cite{Kirillova2016,Campuzano2019,Liu2022}, and in the field of cargo transport\cite{Baraban2012,Demirörs2018,Erez2022}. Some of these Janus particles can self-propel under suitable environmental conditions. The most frequently studied self-propulsion mechanisms are self-diffusiophoresis\cite{Howse2007,Jalilvand2018}, self-electrophoresis\cite{Paxton2004}, and self-thermophoresis\cite{Jiang2010,Bregulla2015,Heidari2020}. Self-diffusiophoretic (e.g., Platinum-PS\cite{Howse2007,Jalilvand2018}) and self-electrophoretic (e.g., Platinum/Au\cite{Paxton2004}) Janus particles rely on external fuel (concentration gradient, self-generated electrical field) to propel themselves. The movement continues until the fuel is consumed. Self-thermophoretic Janus particles, in contrast, convert laser light into heat, which allows the active movement to be controlled by the laser light intensity.\\
A particular example of self-thermophoretic Janus particles are partially Au-coated polystyrene particles (Au-PS). A green laser ($\lambda$ = 532\,nm) is preferentially absorbed by the Au-coated side of the particle, which generates a temperature gradient between the heated Au side and the uncoated PS side. This temperature gradient induces a phoretic flow around the particle that causes the self-propulsion.\cite{Jiang2010,Bickel2013,Bregulla2015,Kroy2016}\\
There are several possibilities to produce the partially Au-coated Janus particles. The most common technique is the plain metal deposition method because of its simplicity. This method is based on the fabrication of a particle monolayer on a solid substrate, which is followed by a coating of the upper part of the particles with a metal layer\cite{Walther20082}. For the metal deposition step, sputter coating\cite{Suzuki2006,Zong2015,Xuan2016,Chen2018} or thermal evaporation\cite{Jiang2010,Buttinoni2012,Bregulla2014,Nedev2015,Simoncelli2016,Feldmann2019,Heidari2020,Heidari2021} are the most commonly used techniques to fabricate Au-coated Janus particles. During sputtering, highly energetic ions strike a solid target (in our case, Au or chromium) and knock atoms out off the surface of the target material\cite{Swann1988}. In a thermal evaporation process, the desired material (Au or chromium) is heated in vacuum until it evaporates\cite{Swann1988}. The Au/chromium atoms then form a thin film on the substrate, which contains the PS particles resulting in partially coated Janus particles. To fabricate Janus particles that are coated less than half, it is possible to embed the particles in a polymer matrix prior to the metal deposition\cite{Paunov2004,Paunov2005} and to separate the partially coated particles afterwards.\\
The effect of particle coverage on the resulting phoretic velocity was studied experimentally by Jalilvand\,\textit{et al.}\cite{Jalilvand2018}. They investigated Pt-coated PS particles undergoing self-diffusio-phoresis in a H$_2$O$_2$ solution. The coated particle area was calculated depending on the patch geometry. They found that a decrease in coverage from 50\,\% to 25\,\% decreased the phoretic velocity by less than 10\,\%. Golestanian\,\textit{et al.}\cite{Golestanian2007} presented general considerations for the design of small phoretic microswimmers. They theoretically described that fast motion depends on a smart surface design to achieve a large surface "activity" and "phoretic mobility.\\
So far, an experimental investigation of the influence of the fabrication method on the resulting Au cap of Janus particles and the resulting thermophoretic behavior is missing. However, this investigation is tremendously important not only to ensure the comparability of different studies but also to maximize the system's efficiency through the interplay of laser power and coverage. Furthermore, nearly every group uses a different device to manufacture their Au-coated Janus particles. Even though in the literature the Au is often deposited directly onto the particles\cite{Suzuki2006,Zong2015,Xuan2016,Chen2018,Jiang2010,Nedev2015,Simoncelli2016}, some groups deposit a few nanometer-thick chromium layer underneath the Au layer to improve the adhesive properties\cite{Heidari2020,Heidari2021,Buttinoni2012,Bregulla2014,Feldmann2019}.\\
In the present study, we investigate the influence of several commonly used preparation methods of Janus particles on the resulting Au cap and the self-propulsion behavior. The Janus particles are analyzed with various techniques and a new method is introduced. Partially Au-coated PS particles are prepared by sputter coating, thermal evaporation, or a combination of particle embedding and additional sputter coating. The impact of different sputter coating devices operating at various sputtering rates and the optional chromium layer on the resulting Janus particle properties is examined. First of all, the Au cap of the Janus particles is analyzed with scanning electron microscopy\cite{Jalilvand2018,Volpe2011,Suzuki2006,Zong2015,Xuan2016,Nedev2015,Ye2010} (SEM) and energy dispersive x-ray\cite{Xuan2016,Ye2010} (EDX) analysis to probe the capping. Additionally, a novel non-invasive method is introduced, which allows a quantitative calculation of the Au cap size by analyzing the shielded substrate areas with atomic force microscopy (AFM) after Janus particle removal. The self-propulsion behavior of the differently prepared Janus particles was examined with a dark-field microscope (DFM), where the measurements are focused on a laser range below 10\,mW. In the discussion, the influence of the preparation technique on the obtained thermophoretic velocity and diffusion constant is evaluated.

\section{2. Experimental Section} \label{Experimental Details}
\subsection{2.1 Materials}
Polystyrene (PS) particles with a diameter of 2.39 µm were purchased from microparticles GmbH (Berlin, Germany) as 10\,\% w/v aqueous suspension. Phytagel$^\text{TM}$ (agar substitute produced from a bacterial substrate composed of glucuronic acid, 
rhamnose, and glucose) was acquired from Sigma Aldrich (Darmstadt, Germany) and Sylgard\textregistered\,184 was supplied from Dow Corning (Midland, USA). Deionized water (resistance 18.2\,M\textOmega\,cm at 25\,$^\circ$C) was obtained from a MilliQ water purification system from Merck (Darmstadt, Germany). Microscope slides (LABSOLUTE, pure white, 76\,mm\,$\times$\,26\,mm\,$\times$\,1\,mm) and MgSO$_4$ (Chemsolute) were purchased from Th. Geyer GmbH + Co. KG (Renningen, Germany). Microscope cover glasses (18\,mm\,$\times$\,18\,mm) were acquired from Paul Marienfeld GmbH + Co. KG (Lauda-Königshofen, Germany). Silicon paste (Baysilone-Paste) and 2-Propanol ($\geq$ 99.5\%, for synthesis) were obtained from Carl Roth GmbH + Co. KG (Karlsruhe, Germany). Ethanol (EtOH) was purchased from Fisher Chemical (Waltham, USA). AC160TS-R3 (300\,kHz, 26\,N/m) cantilevers were purchased from Oxford Instruments (Abingdon, UK). \\

\subsection{2.2 Au-PS Particle Preparation} \label{sec:Au-PS particle preparation}
\subsubsection{Sputtering/Thermal Evaporation} \label{sec:SputteringThermal evaporation}
\begin{figure}[H]
\centering
\includegraphics[scale=1.2]{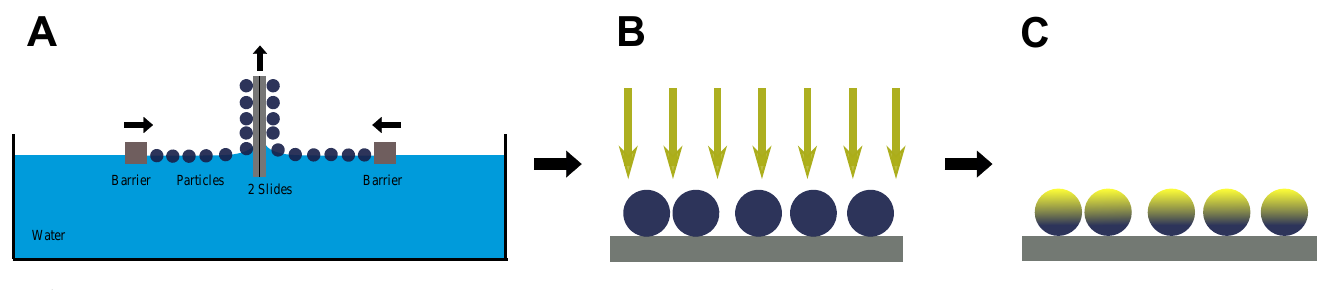}
\caption{Schematic of the Janus particle preparation process via sputter coating or thermal evaporation. Part A: Transfer of PS particles on two microscope slides via the Langmuir-Blodgett technique, part B: Metal deposition step, and part C: Au-PS Janus particles.}
\label{fig:Sketch1}
\end{figure}

Prior to use, the standard microscope slides were cleaned for 30\,min in ethanol in an ultrasonic bath. Two microscope slides were then taped together using double-sided adhesive tape and thus a monolayer of particles could be transferred to both slides simultaneously via the Langmuir-Blodgett technique\cite{Peterson1990,Lu2006} (see Figure\,\ref{fig:Sketch1}, part A). The two microscope slides were then carefully detached from each other. After drying in ambient conditions, the wafers were placed in the corresponding physical vapor deposition system to coat the particles with a thin layer of Au (see Figure\,\ref{fig:Sketch1}, parts B and C). Optionally, an additional layer of chromium was deposited prior to the Au layer under the same conditions to enhance the attachment of Au to the microscope slides. Thus, a potential influence of the chromium layer could be investigated.

\noindent Au was either sputter coated or thermally evaporated.\\
\textit{Sputter coating}: The monolayer of PS particles was sputter-coated with an Edwards Sputter Coater S150B, a BOC Edwards Auto 500, or a Q300TD Sample Preparation System (Quorum Technologies, UK).\\
\textit{Thermal evaporation}: For thermal evaporation, a CREAMET 300 V2 evaporator was used. \\ \\
Table\,\ref{SI-table:Parameters} in the SI summarizes the parameters of the coating process, such as the rates of the sputtering or thermal evaporation processes, the corresponding times, the adjusted values for current or power, and the distance of the substrate to the evaporation source.\\
The thickness of the resulting metal layer during the fabrication process was estimated using a reference crystal from an in situ quartz crystal microbalance (QCM-D) that was coated simultaneously and mounted in the coating devices. After the deposition process, the prepared Janus particles can be simply removed from the substrate with a few water droplets. As a result, the dispersion of Janus particles in water is obtained. When an additional chromium layer was added in the preparation process, the excess Au from the glass substrate adheres well enough to the glass that only the coated particles are removed. For the samples without chromium, an additional filtering step (using a paper filter with a pore size $>$ particle diameter) becomes necessary to remove Au flakes, detached from the glass slide.

\subsubsection{Gel Trapping Technique}
\begin{figure}[H]
\centering
\includegraphics[scale=1.3]{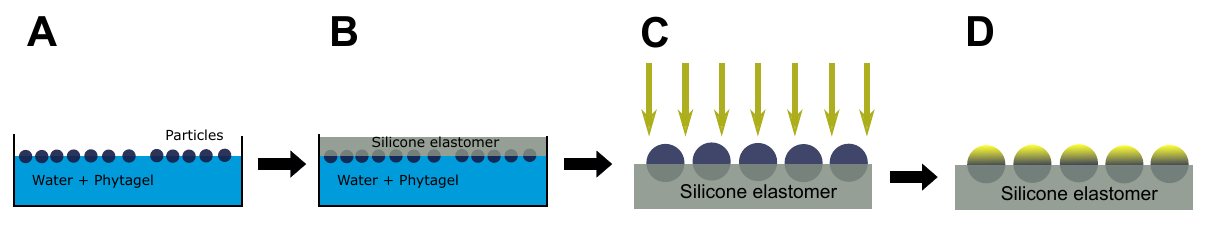}
\caption{Schematic of the gel trapping technique. Part A: Trapping of PS particles at the interface of the Phytagel solution, part B: Pouring of liquid silicone elastomer on top of trapped PS particles + curing, part C: Metal deposition on embedded PS particles, and part D: Au-PS Janus particles in the matrix.}
\label{fig:Sketch2}
\end{figure}

To prepare Janus particles with smaller Au caps, one has to partially embed them in a matrix of silicone elastomer. To achieve this, a gel trapping technique similar to that described by Paunov and Cayre\cite{Paunov2004,Paunov2005} was applied. To prepare the PS particle suspension, 0.167\,ml of the commercial 10\,\% w/v aqueous PS particle suspension was diluted with 0.333\,ml MilliQ water. This was then mixed with 1\,ml EtOH.\\ 
Phytagel$^\text{TM}$ powder (see concentrations in Table\,\ref{SI-table:Phytagel} in the SI) was dissolved in MilliQ water by continuous stirring at 85\,$^\circ$C. The hot Phytagel solution was filled in a petri dish. After two minutes, the PS particle solution (about ten to fifteen droplets of the prepared PS particle suspension) was spread with a pipette directly onto the Phytagel solution (see Figure\,\ref{fig:Sketch2}, part A). Cooling down to room temperature leads to subsequent gelation of the Phytagel/water phase and as a result, the PS particles were - partly sunken into the gel - directly trapped at the interface. For the variation of the Au cap size of the Janus particles, prior to sputtering the penetration depth of the particles into the Phytagel was varied. To achieve different degrees of embedding, the Phytagel concentration (Em1: 0.1\,g in 10\,ml water; Em2: 0.01\,g in 10\,ml water) and the divalent cation concentration (Em1: without; Em2: 4\,mM MgSO$_4$) was varied between two batches. According to the supplier of Phytagel (Sigma Aldrich), in low-salt media formulations (such as MilliQ water), one needs either a higher concentration of Phytagel or supplementation with magnesium or calcium salts (e.g., CaCl$_2$ or MgSO$_4$) to form a gel. A sample without additional divalent cations and a Phytagel concentration below 0.1\,g in 10\,ml water did not undergo gelation and therefore could not be used for the production of Janus particles. The addition of divalent ions facilitates the Phytagel (composed of glucuronic acid, glucose, and rhamnose monomers) gelation due to its complexation with the carboxylate moieties.\cite{Loosli2016} This enhanced gelation ability leads to less deep embedding of the Em2 particles (with divalent ions) into the Phytagel and deeper embedding into the silicone elastomer layer compared to the Em1 particles (without divalent ions). The absence of divalent ions prevented Phytagel solution from fully gelating. By applying the liquid silicone elastomer to the polystyrene particles trapped at the interface, the Em1 particle then further sink into the Phytagel layer and thus are less deeply embedded in the elastomer matrix than the Em2 particles.\\
The last batch (Em3) was prepared without Phytagel solution and the silicone elastomer was poured directly onto the PS particle monolayer on a microscope slide. This was sufficient here since only different degrees of embedding were to be achieved.\\
Within a batch (with a given Phytagel concentration, see Table\,\ref{SI-table:Phytagel} in the SI and the text there), the degree of embedding of PS particles in the matrix barely varied. To further modify the degree of embedding, for instance, different oils could be used as a second phase (instead of air)\cite{Cayre2004}. Further information about the gel trapping technique can be found in the SI in chapter 1. Immediately after evaporation of EtOH from the spread PS suspension, a liquid silicone elastomer (Sylgard\textregistered\,184, Dow Corning) was poured on top of the trapped particles and cured for at least 48\,h at room temperature (see Figure\,\ref{fig:Sketch2}, part B). After the curing step, the silicone elastomer could be carefully peeled off, and the lower part of the particles was trapped inside the silicone elastomer (see Figure\,\ref{fig:Sketch2}, part C) as evidenced by scanning electron microscopy (see Figure\,\ref{fig:SEM} (d)\,-\,(f)). Afterwards, they were coated with Au as described above (see Figure\,\ref{fig:Sketch2}, part D).

\noindent Further information including the amount of Phytagel, the release of the Janus particles from the matrix, and modifications of the embedding process can be found in the SI in chapter 1.\newline
Table\,\ref{tab:OverviewJanus} summarizes all studied Janus particles with details about their preparation.

\begin{table} [H]
\caption{Overview of Janus particles analyzed in this paper, including the preparation method, and the aimed cap thickness. More details can be found in the SI.}
\centering
\begin{tabular}{lcl}
\hline
Sample           & Method & Aimed cap thickness  \\
\hline    
Sc1  & \multirow{5}{*}{Sputter coating} &  50 nm Au     \\
Sc2  &  & 50 nm Au \\
Sc3  &  & 5 nm Cr + 50 nm Au  \\
Sc4  &  & 50 nm Cr + 50 nm Au \\
Sc5  &  & 5 nm Cr + 50 nm Au \\
&&\\
Th1  & Thermal evap. & 5 nm Cr + 50 nm Au \\
&&\\
Em1  & \multirow{1}{*}{Embedding} & 50 nm Au \\
Em2  & \multirow{1}{*}{ + } & 50 nm Au  \\
Em3  & \multirow{1}{*}{Sputter coating} & 50 nm Au \\
\hline
\multicolumn{3}{c}{}
\end{tabular}
\label{tab:OverviewJanus}
\end{table}

\subsection{2.3 Characterization Techniques} \label{subsec:charactTechniques}

\textbf{SEM:} Scanning electron microscopy of the embedded particles and finally the Janus particles was carried out using a Gemini 500 Field Emission Scanning Electron Microscope (FESEM) from Carl Zeiss (Oberkochen, Germany).  With this FESEM, also elemental analysis by using energy dispersive x-ray spectroscopy (EDX; Quantax EDS from Bruker, Massachusetts, USA) is possible and was used to obtain further information on the Au cap of the Janus particles. Note, that uncoated “shadows” (= "footprint") form underneath the spheres during the process of sputtering/thermal evaporation, where the particles have shielded the substrate from the incoming Au.\\

\noindent\textbf{AFM:} Topographical images of the substrate after removal of the prepared Janus particles were carried out in tapping mode at ambient conditions on a MFP3D SA (Asylum Research/Oxford Instruments, Wiesbaden, Germany). These topographical images of the "footprint" of Janus particles on the substrate serves to determine the thickness and area of the Au cap. This novel method is described in the result section. As a cantilever, AC160TS-R3 (300 kHz, 26 N/m) was used. The uncoated areas and their transition heights were analyzed with the AFM built-in software features based on IGOR 16.\\

\begin{figure}[H]
\centering
\includegraphics[scale=0.55]{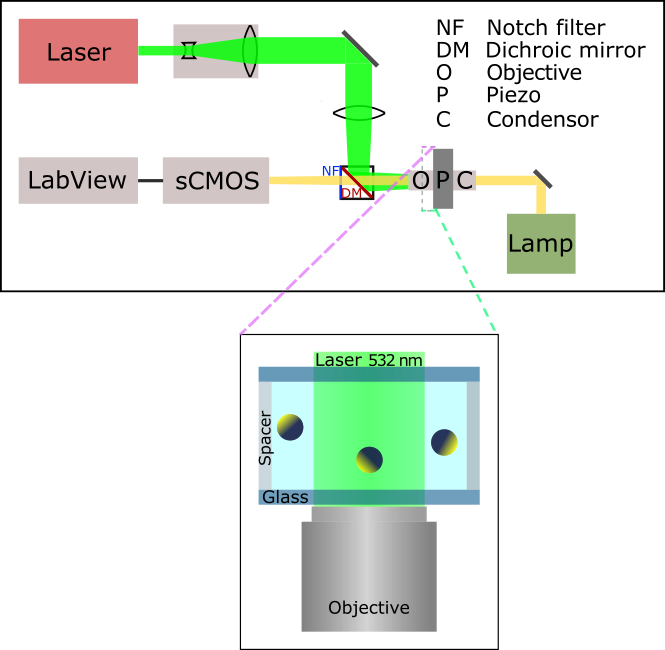}
\caption{Dark-field microscopy setup with an enlarged view of the sample cell. The laser beam propagates through a beam expander and a lens subsequently to end up in a wide parallel beam at the sample location. The sample cell is fixed on a nano-positioning stage on top of the objective and a camera is used to image the Janus particle.}
\label{fig:DFM}
\end{figure}

\noindent\textbf{Self-Propulsion Measurements:} The sample cell (see Figure\,\ref{fig:DFM}, inset) for the self-propulsion measurements consists of two bare glass slides, which were pre-cleaned for 30 min in ethanol in an ultrasonic bath. Afterwards, a 1\,\textmu l droplet of the dispersion of Au-PS particles is deposited between those glass slides (18\,$\times$\,18\,mm$^2$). To restrict the lateral movement of the particles and prevent evaporation of water inside the sample cell, the edges of the glass slides were covered with silicon paste. The sample cell is positioned on top of the objective and by coupling the laser into the system, self-propulsion of the Janus particles is observed.\\
The self-propulsion measurements are conducted at a dark-field microscope set-up (see schematically in Figure\,\ref{fig:DFM}) increasing the sensitivity to the Au cap (condenser, Olympus, NA 1.2-1.4). An oil-immersion objective (Olympus 100x, adjustable NA 0.5-1.35) collects the scattered light of the particles in the sample cell and with an sCMOS camera (Andor, Zyla 4.2), this scattered light is imaged. A green laser (Pegasus, Pluto, 800\,mW, 532\,nm) is coupled to the microscope and stimulates the self-propulsion of the Au-PS particles. With a LabView program, the x- and y-position of the analyzed particle in every frame is recorded in real-time. A nano-positioning stage is used to automatically adjust its position so that the particle is again at the center of the illumination. Further set-up details can be found in Heidari \textit{et al.}\cite{Heidari2020,Heidari2021}. \\
For lateral movement at times well below the rotational diffusion time $\tau_\text{R} = (8 \pi \eta R^3)/(k_\text{B} T) = 1/D_\text{r} \approx$ 10\,s (viscosity of water $\eta$, particle radius $R$, and thermal energy $k_\text{B} T$) of a particle (t $<<$ $\tau_\text{R}$), the mean square displacement (MSD) can be described by\cite{Uhlenbeck1930,Howse2007,Martens2012}:
\begin{align}
    \text{MSD} = 4 D t + v_{\text{th}}^2 t^2,
    \label{eq:MSD}
\end{align}
where $v_{\text{th}}$ is the thermophoretic velocity of a particle and $t$ is the lag time. By fitting the obtained MSD curves with equation\,\ref{eq:MSD}, the diffusion coefficients $D$ (see Figure\,\ref{SI-fig:Diffusion_coefficient}) and the thermophoretic velocities $v_{\text{th}}$ (see Figures\,\ref{fig:vth} and \ref{SI-fig:vth_SI}) of all particles can be extracted. The error bars represent the standard deviations of at least 50 measurement values for each data point.

\newpage

\section{3. Results} \label{Results}
We have used three different methods (sputter coating, thermal evaporation, and a combination of particle embedding and subsequent sputter coating) to prepare partially Au-coated Janus particles. First, the Janus particles are characterized by SEM, EDX analysis, and AFM measurements. Second, self-propulsion measurements at different laser powers are carried out using a dark-field microscope set-up with a nano-positioning stage. Thereby, the influence of the sputter coater machine/sputter rate, an underlying chromium layer, and most important the Au cap size on the resulting thermophoretic velocity is investigated.

\subsection{3.1 Janus Particle Characterization}\label{sec:Janus particle characterization}
\begin{figure}[H]
\centering
\includegraphics[scale=0.40]{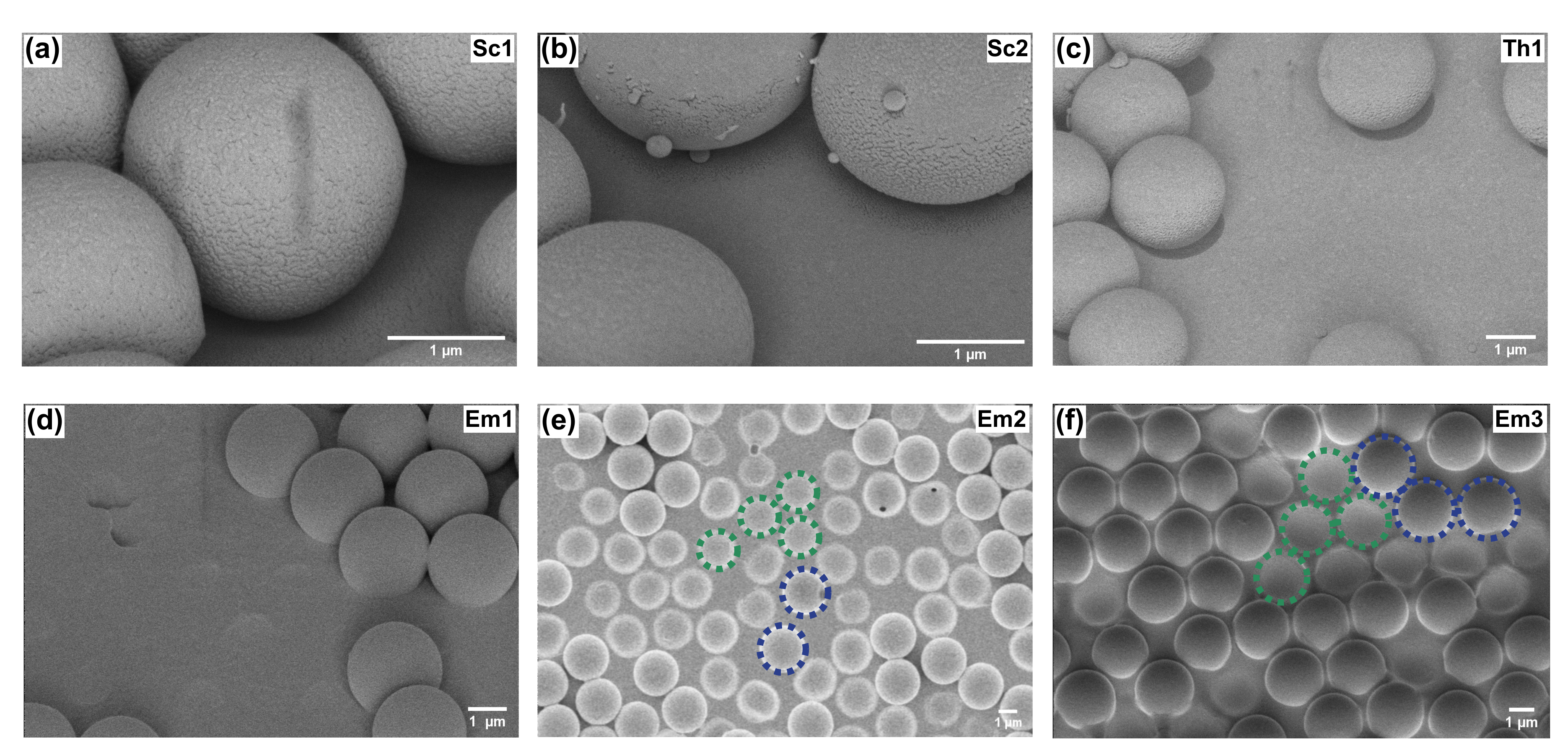}
\caption{SEM images of (a) and (b) Janus particles Sc1 and Sc2 obtained by sputter coating, and (c) Th1 particles obtained by thermal evaporation. SEM images of embedded (d) Em1, (e) Em2, and (f) Em3 particles, where the green circles in (e) and (f) correspond to holes in the matrix (where once a particle was located) and the blue circles indicate particles that are still embedded in the matrix. The degree of embedding increases from left to right. The white error bar always represents 1\,\textmu m.}
\label{fig:SEM}
\end{figure}

Selected SEM images of the Janus particles after sputter coating and thermal evaporation are shown in Figure\,\ref{fig:SEM}\,(a)-(c). Figure\,\ref{fig:SEM}\,(d)-(f) shows the particles partially embedded in the silicone elastomer matrix with varying degrees of embedding after peeling off the elastomer from the Phytagel. Additional SEM images of the remaining Janus particles and, for comparison, an uncoated PS particle can be found in the Supporting Information (Figure\,\ref{SI-fig:Pure_PS} and Figure\,\ref{SI-fig:SEM_SI}).\\
As shown in the following, the Au coating is thickest at the top of the particle which was oriented towards the Au source. This part is called the "pole" in the following, which defines also the equatorial plane.

\noindent\textit{SEM}: For the particles obtained by sputter coating (Figure\,\ref{fig:SEM}\,(a) and (b)), a change in coating properties from the particle pole to the particle equator is observed. But, the exact line, where the Au coating ends is not clearly discernible from the simple SEM scans right after the preparation process. On the substrate, a black shadow is visible around the spot where the particles are adsorbed at the substrate. This represents the area that was not coated with Au due to shielding by the particle. This is analyzed more in detail in section\,3.2. In the SEM image of the particles obtained by thermal evaporation (Figure\,\ref{fig:SEM}\,(c)), this black shadow below the Janus particles is noticeably larger than for the sputter-coated particles. The lower regions of those particles appear to be uncoated, but still, the exact boundary between the Au cap and the pure PS is not discernible.\\
For the particles embedded in the silicone elastomer matrix, it can be clearly seen that the intended difference in protruding height was achieved (see Figure\,\ref{fig:SEM}\,(d)\,-\,(f)). The particles of a respective batch Em1-Em3 are embedded to a similar depth in each case. The hole diameter in the silicone elastomer matrix after particle removal can be found in Table\,\ref{SI-table:Embedding}. This diameter varies between 2.8\,\% (Em3) and 11.2\,\% (Em1). However, the Em1 particles were embedded so shallowly in the silicone elastomer matrix that the part coated by sputtering completely peeked out of the matrix. Thus, it does not matter how differently deep (11.2\% variation)  the individual particles of this batch were embedded.

\begin{figure}[H]
\centering
\includegraphics[scale=0.40]{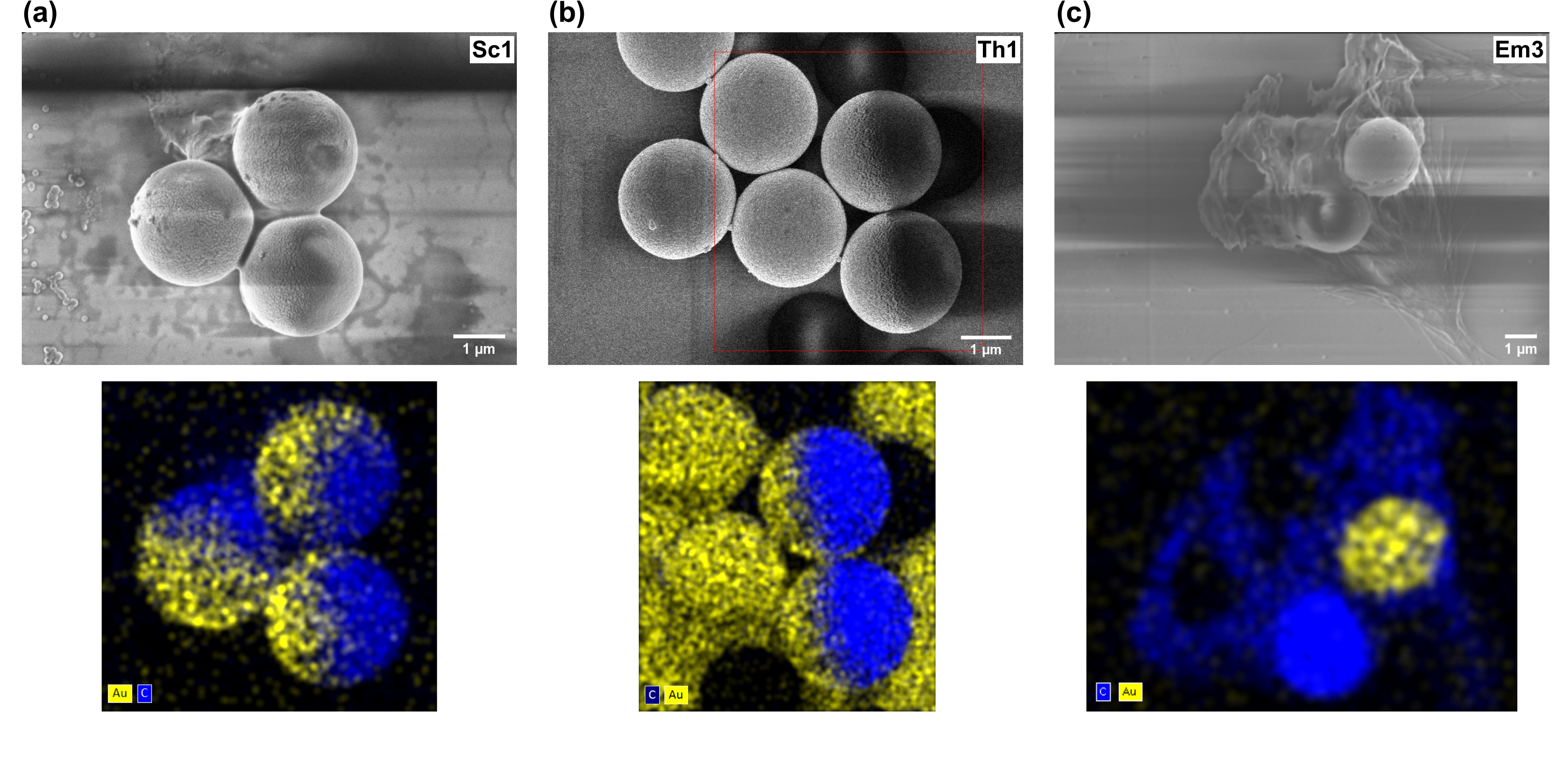}
\caption{SEM images (top row) with corresponding results of the EDX analysis (bottom row) of (a) particles Sc1 obtained by sputter coating, (b) particles Th1 obtained by thermal evaporation, and (c) embedded particles Em3, where the yellow part corresponds to the Au cap and the blue part to the PS side.}
\label{fig:EDX}
\end{figure}

\noindent\textit{EDX analyses:} To demonstrate genuinely that the particles were partially coated with Au, additional EDX analyses were performed that allowed mapping of the polystyrene and the Au side. Figure\,\ref{fig:EDX} shows the EDX analyses of Janus particles with an additional SEM image of the corresponding area. The blue part in the EDX images corresponds to the uncoated polystyrene side and the yellow part represents the Au coating. Further EDX analyses of the remaining particles and SEM images with a different detector can be found in the Supporting Information in Figures\,\ref{SI-fig:EDX_SI_1}\,-\,\ref{SI-fig:SEM_SI_Detector_2}. Additionally, the boundary between the Au cap and the uncoated PS part of some particles (see Figure\,\ref{fig:EDX}\,(a) and Figure\,\ref{SI-fig:SEM_SI_Detector_2}\,(a) and (b)) is not a straight line but shows a wavy pattern.\\
Even though the Au cap can be detected well in the EDX images, quantitative analyses are very complex due to the limited spatial resolving capability of EDX. Additionally, to see the transition from Au to PS accurately, the poles of the particle preferably have to lie in the imaging plane but the particles are randomly oriented after their detachment from the wafer. Consequently, these EDX analyses are considered to be only qualitative. Therefore, a new and also non-invasive method was found that enables quantitative analysis of the Au cap which will be introduced in more detail in the next section.

\subsection{3.2 Novel AFM-based technique for Au cap size determination} \label{subsec:Novel technique}
\begin{figure}[H]
\centering
\includegraphics[scale=0.57]{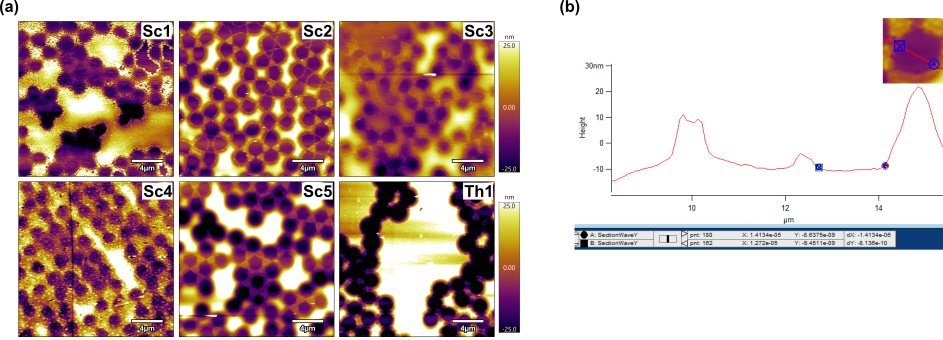}
\caption{AFM images of the residual Au layer on the substrate after particle removal of the non-embedded Janus particles (a) and analysis method with the AFM built-in features where a line is drawn through a single particle and the diameter of the dark area is analyzed (b). The inset shows the corresponding area with the two markers.}
\label{fig:Shadow}
\end{figure}

To quantitatively investigate the Au cap size, atomic force microscopy measurements of the "footprints" of the particles were carried out: After removal of the particles, the microscope slide was scanned by AFM. From the size of the holes in the Au layer, which were left by the particles, the size of the Au cap can be determined. The thickness of the Au cap at the upper pole is determined from the thickness of the Au film at the substrate where the transition height from the uncoated "footprint" to the gold film was scanned at suitable points. Figure\,\ref{fig:Shadow}\,(a) shows the measured topography of the residual Au layer after removing the particles. In Figure\,\ref{fig:Shadow}\,(b), an analysis line within the AFM software is shown, where the distance between the two markers corresponds to the diameter of the uncoated/shadowed area. 

\begin{figure}[H]
\centering
\includegraphics[scale=2.6]{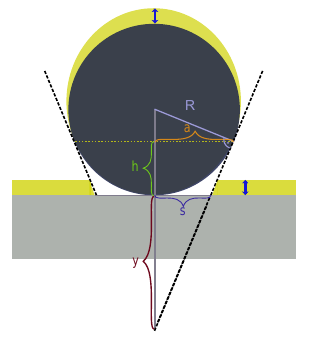}
\caption{Illustration of a single Janus particle with radius $R$ after the coating process: The particle is partially covered with a thin Au cap, which has its maximum thickness on the top (pole) of the sphere and becomes thinner towards the particle equator\cite{Rashidi2018}. Additionally, during the coating process the particle has shielded an area with radius $s$ ("footprint").}
\label{SI-fig:Model}
\end{figure}

\noindent The unknown variable we want to determine using the equations provided below is the height of the uncoated part $h$ because from this we can first calculate the uncoated surface area $O_\text{R,h}$ (PS side) of the Janus particle:

\begin{align}
O_\text{R,h} = 2 \pi R h
\end{align}

Depending on the depicted situation in Figure\,\ref{SI-fig:Model}, three main equations can be derived:
\begin{align}
\frac{s}{a} & = \frac{y}{y+h}\\
R^2 & = (R-h)^2 + a^2 & \\
a^2 & = (R-h) \times (h+y)
\end{align}

\noindent The particle radius $R$ is known from the manufacturer of the PS particles and $s$ can be measured with an AFM as the radius of the uncoated area as described above.

\noindent Solving the equations (1) - (3) one obtains for $h$: 

\begin{align}
h = \frac{2 R s^2}{R^2 + s^2}
\end{align}

\noindent The total particle surface area is $O_\text{R} = 4 \pi R^2$. Consequently, the size of the Au cap, i.e., the degree of coverage of the particle surface with Au as a function of the particle radius $R$ and radius of the shielded uncoated area under the particle $s$ can be calculated as the difference of total surface area $O_\text{R}$ and uncoated surface area $O_\text{R,h}$:

\begin{align}
O_\text{R, gold cap} & = O_\text{R} - O_\text{R,h} = 2 \pi R (2 R - h)
\end{align}

\begin{figure}[H]
\centering
\includegraphics[scale=0.75]{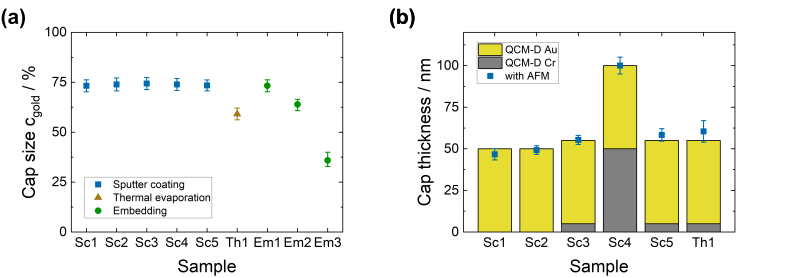}
\caption{(a) Au cap size of all investigated samples deduced from the uncoated/shadowed area or for the embedded particles from SEM images. The dark blue squares correspond to the Janus particles obtained by sputter coating, the brown triangle to the particles obtained with thermal evaporation, and the green circles to the embedded and additionally sputter-coated particles. (b) Comparison of cap thickness determined via QCM-D (Cr cap + Au cap thickness) during the preparation process (grey + yellow bars) and cap thickness measured with AFM (blue squares).}
\label{fig:Gold_cap}
\end{figure}

\noindent The Au cap size $c_{\text{gold}}$ (in \%) in Figures\,\ref{fig:Gold_cap}\,(a) and \ref{fig:vth} of this study refers to the relative coverage of the particle with Au, i.e., what percentage of the surface of the PS particle is covered with Au. This is obtained by the quotient of the Au cap size and the total particle surface:

\begin{align}
c_{\text{gold}} = \frac{O_\text{R, gold cap}}{O_\text{R}} = \frac{2 \pi R (2 R - h)}{4 \pi R^2} = \frac{2 R - h}{2 R}
\end{align}

\noindent A detailed description of the evaluation of the Au cap size $c_{\text{gold}}$ of the embedded particles can be found in the SI in chapter 2.\\ \\

\noindent The calculated Au cap sizes of the Janus particles obtained by sputter coating and thermal evaporation are displayed in Figure\,\ref{fig:Gold_cap}\,(a) (blue squares and brown triangle). Regardless of the sputter type or sputter rate or the underlying chromium layer (0 nm, 5 nm, or 50 nm), the Au cap size $c_{\text{gold}}$ of all Janus particles obtained by sputter coating is approximately 74\,\%. Compared to this, thermal evaporation results in smaller values for $c_{\text{gold}}$ of about 59\,\%.\\
For the determination of $c_{\text{gold}}$ of the embedded particles (green circles in Figure\,\ref{fig:Gold_cap}\,(a)), the Au cap size can be determined by analyzing the holes in the matrix after particle removal (see SI, chapter 2) or the part of the particles which is not embedded in the silicone elastomer. Due to a varying degree of embedding, variable Au cap sizes can be achieved, ranging from about 74\,\% to 64\,\% to values of about 36\,\%, where consequently less than half of a particle is coated with Au.\\  
Figure\,\ref{fig:Gold_cap}\,(b) displays the maximum cap thicknesses determined by AFM compared to those determined by QCM-D (in the coating cell). The thickness values obtained by both methods are in good agreement.\\

\subsection{3.3 Thermophoretic Self-Propulsion}
\begin{figure}[H]
\centering
\includegraphics[scale=0.50]{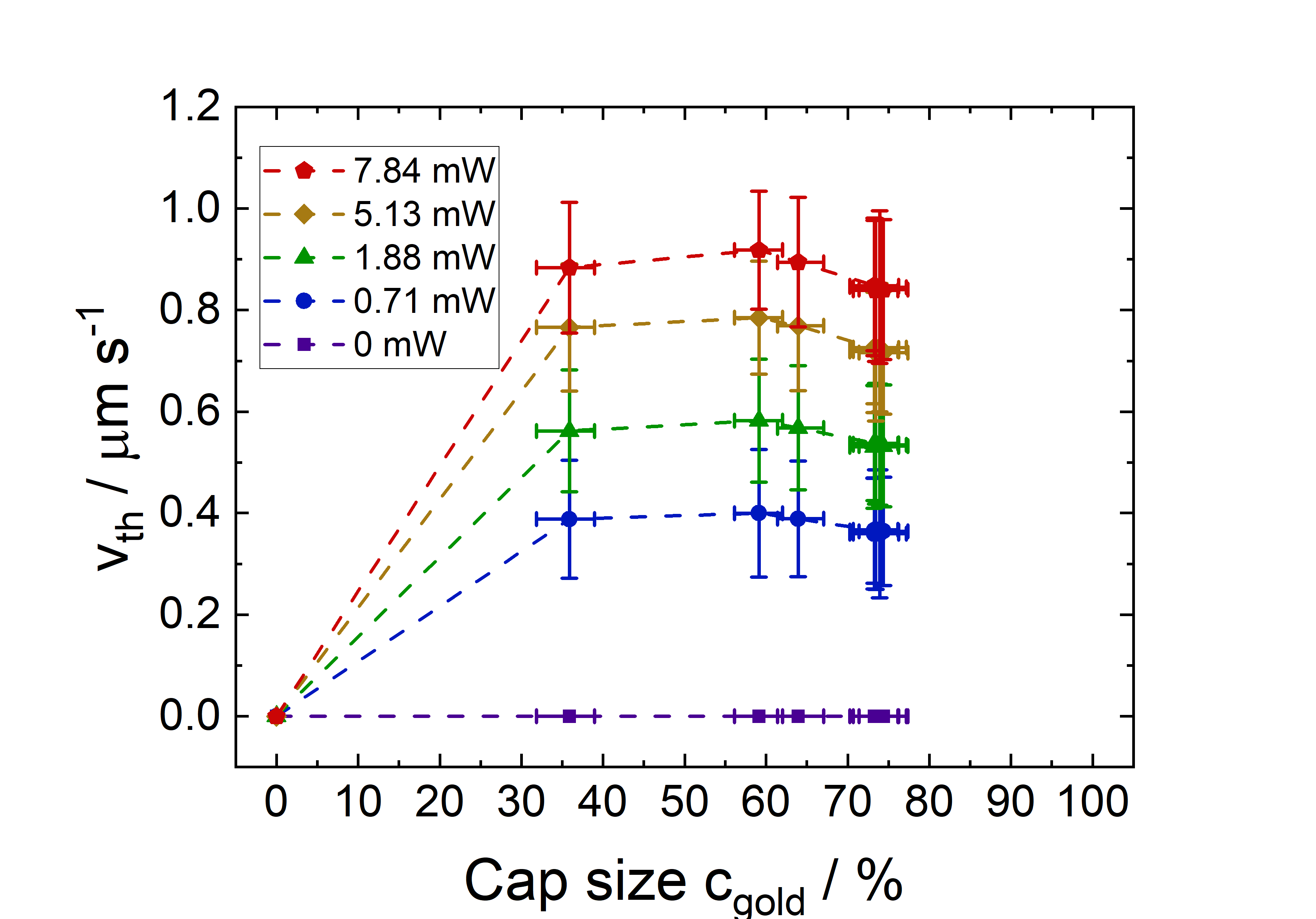}
\caption{Thermophoretic velocity in dependence of the Au cap size for different laser powers.}
\label{fig:vth}
\end{figure}

Figure\,\ref{fig:vth} presents an overview of the examined parameters that potentially alter the thermophoretic velocity of the Janus particles. The corresponding mean square displacement (MSD) curves and trajectories as well as the diffusion coefficients $D$ are provided in the SI Figures\,\ref{SI-fig:Diffusion_coefficient} - \ref{SI-fig:MSD_traj2}. Figure\,\ref{fig:vth} shows the impact of the Au cap size $c_{\text{gold}}$ on the resulting thermophoretic velocity $v_{\text{th}}$. The influence of the underlying chromium layer and the sputter rate on $v_{\text{th}}$ are summarized in Figures\,\ref{SI-fig:vth_SI}\,(a) and (b) in the SI.

\noindent Neither a 5\,nm thin chromium layer nor the sputter rate of the different sputter devices show any discernible influence on the resulting thermophoretic velocity. Only the 50\,nm thick chromium layer leads to a slight decrease in $v_{\text{th}}$ compared to the corresponding values without or a 5\,nm thin chromium layer.\\
For a particle whose surface has no asymmetry (uncoated particle with 0\,\% Au cap), the thermophoretic velocity vanishes for all laser powers as expected. When the Au cap covers between about 36\,\% and 74\,\% of the particle surface, the particle shows self-propulsion with a similar velocity $v_{\text{th}}$ at a fixed laser power. The effect of cap size in this cap size regime (36\,\% and 74\,\%) is very low if there is any. $v_{\text{th}}$ exhibits the highest values for a 59\,\% Au cap and slightly decreases for the here analyzed larger (64\,\% and 74\,\%) and smaller (36\,\%) Au caps, but the differences vanish in the error bars. However, the plot of $v_{\text{th}}$ suggests that for smaller ($<$ 36\,\%) Au caps, the influence of the Au cap size on the velocity might increase as the velocities vanish at this extreme value. Assuming also a vanishing $v_{\text{th}}$ for a fully coated Janus particle with a Au cap size of 100\,\%, one could suppose that the influence of the Au cap also becomes more pronounced for values of $c_{\text{gold}} > $ 74\,\%.

\newpage
\section{4. Discussion}\label{Discussion}
\subsection{4.1 Au cap characterization}
Au-PS Janus particles were successfully fabricated with varying methods and characterized regarding their Au cap and self-propulsion behavior. The Janus particles obtained with sputter coating have a larger Au cap compared to the ones after thermal evaporation (see Figure\,\ref{fig:Gold_cap}\,(a)). This could be explained by the fact that sputtered atoms have a higher energy than thermally evaporated ones\cite{Swann1988,Golan1991,Bendavid1999}. Consequently, the sputtered Au atoms can penetrate further and at different angles (due to more collisions) into the polystyrene particle monolayer during the coating process compared to the thermally evaporated ones. Thus, the sputtered Au atoms could cover a larger part of the PS particles at the lower polystyrene particle hemisphere below the equator. Both, sputtering and thermal vapor deposition result in Janus particles that are slightly more than half-coated. However, the Janus particles fabricated by thermal vapor deposition with a Au cap size of about 59\,\% are approximately half-coated. By embedding the particles in a silicone elastomer matrix, Janus particles covered with less than 50\,\% Au could be prepared. This is consistent with observations of Paunov and Cayre\cite{Paunov2004,Paunov2005} and Adams \textit{et al.}\cite{Adams2008}, who coated their particles after an embedding step with Au. The SEM images of their Janus particles indicate that less than 50\,\% of each particle is coated with Au.\\
A tight packing in hcp order of the PS particles shields these areas of the particles from getting coated, which results in a wavy boundary line between the Au and the PS side (e.g., see Figure\,\ref{fig:EDX}\,(a)). This wave-like boundary layer between coated and uncoated part was also observed by Zong \textit{et al.}\cite{Zong2015} after sputtering a hexagonal-close packed particle monolayer. Increasing the distance between the particles or particle embedding leads to a more even boundary line between the Au cap and the PS side (see Figures\,\ref{fig:EDX}\,(b) and (c)).\\
In the following, the novel technique for Au cap size determination is discussed. Since the lateral resolution of the EDX is low\cite{Burgess2017}, a quantitative analysis of the Au cap based on EDX would be difficult. The method described here assumes a Au cap with a decreasing thickness towards the equator and beyond and not one with a homogeneous thickness because of the particle's curvature. This is consistent with the findings of Rashidi \textit{et al.}\cite{Rashidi2018}, who experimentally analyzed the nominal Au cap thickness of 5\,\textmu m-sized PS Janus particles by using a focused ion beam to cut away a part of the Janus particle and reveal its cross-section. Their measurements have shown that the Au cap thickness is not constant but decreases along the particle's contour towards the equator. A particle shields a certain area (with radius $s$) of the substrate during the coating process, leaving an uncoated area underneath itself. The technique is limited to the quality of the AFM scans and it is sensitive to the fact that more material was detached during the particle removal than expected. As result one obtains a maximum area that is coated with the corresponding coating material (here Au). However, this technique allows an averaging over many particles very easily, because even a single scan area (here 20\,\textmu m\,$\times$20\,\textmu m) contains a large number of residual areas. Another advantage is, that for the measurements here one do not need the Janus particles directly, but only the substrate, which is also an advantage if the number of the produced Janus particles is limited. Additionally, it is also applicable if no SEM is available but any device that is able to resolve the shielded areas. So, this method provides a simple, non-invasive tool to quickly determine the degree of coverage of a particle with Au or any other material that can be imaged.

\subsection{4.2 Thermophoretic Self-Propulsion}
The data show that the preparation technique has almost no effect on the resulting thermophoretic mobility. This is true for the sputter coating devices (with different sputter rates) and for a 5\,nm thick chromium layer. This makes measurements of several groups with various sputtering devices more comparable. Also, it does not seem to matter for the Janus particle properties studied here whether a thin chromium layer is used or not. But an additional thin chromium layer that is added during the coating process simplifies the detachment process of the Janus particles from the substrate. Without this chromium layer, an additional filtering step becomes necessary as Au flakes detach from the glass substrate and have to be filtered out. On the polystyrene particles, on the other hand, the gold seems to adhere, as indicated by the EDX measurements (e.g., Figure\,\ref{fig:EDX} and Figures\,\ref{SI-fig:EDX_SI_1}\,\&\,\ref{SI-fig:EDX_SI_2}). This is in agreement with findings of Himmelhaus and Takei\cite{Himmelhaus2000}, Lahav \textit{et al.}\cite{Lahav2004}, and Zugra \textit{et al.}\cite{Zugra2010} which observed a strong adhesion of Au films on a polystyrene substrate compared to a poor adhesion on a glass substrate in different solvents. Suzuki \textit{et al.}\cite{Suzuki2006} used a polystyrene substrate (instead of a glass substrate) to prepare their Au-coated Janus particles. After particle removal by sonication, the gold still adhered to the polystyrene substrate.\cite{Suzuki2006} To improve the adhesion of Au films on glass substrates, intermediate layers such as chromium can be used.\cite{Pulker1981} Due to oxidation, the chromium adheres very well to the substrate, and the chromium and gold form a diffusion interface layer which also adheres very well.\cite{Pulker1981} Also, e.g., Auschra \textit{et al.}\cite{Auschra2021} added a thin chromium layer before coating of their Janus particles with Au to ensure the adhesion of the Au layer to the glass slide when removing the particles via sonification.\\
In theory, the maximum asymmetry of a Janus particle is reached when exactly half of its surface is coated with Au. At this coverage, one would expect to find the highest thermophoretic velocity. However, this study shows that the thermophoretic velocity only slightly deviates for all analyzed Au cap sizes (see Figure\,\ref{fig:vth}\,(c)). These findings are similar to findings by Jalilvand\,\textit{et al.}\cite{Jalilvand2018} who also investigated the influence of the cap size on the resulting phoretic velocity for Pt-coated PS particles, which undergo self-diffusiophoresis in a H$_2$O$_2$ solution. They also found only a small deviation of the phoretic velocity, which changes less than 10\,\% when varying the cap size from 50\,\% to 25\,\%.\\

\section{5. Conclusion}
In this study, we examine the impact of several commonly used preparation techniques of Janus particles on the resulting Au cap and the self-propulsion behavior. Furthermore, a novel, simple and non-invasive method is introduced, which allows a quantitative determination of the Au cap size.
We successfully fabricated Au-coated PS Janus particles via sputter coating, thermal evaporation, or a combination of particle embedding and a subsequent sputter coating. The Au cap was imaged with SEM and additional EDX analysis and quantitatively investigated with a new technique that based on measuring the uncoated shielded areas underneath the particles. The Au cap size of the particles is influenced by the preparation technique, where the higher energetic Au atoms of the sputtering process cover a larger part of the particle than the less energetic thermal evaporated Au atoms. By partly embedding the particles in a silicone elastomer matrix, even smaller Au cap sizes ($<$\,50\,\%) can be achieved. Neither an underlying chromium layer of 5\,nm or 50\,nm thickness underneath the 50\,nm thick Au cap nor the sputter rate affect the resulting Au cap size. An additional chromium layer of a few nm does not influence the self-propulsion behavior of the particles. Varying the Au cap size in the range of 36\,\% - 74\,\% by the fabrication method has little effect on the resulting velocity of the particles, which simplifies the fabrication process and thus makes data from groups using different manufacturing methods more comparable. The presented data prove that a large variety of partly complementary preparation strategies is applicable to produce Janus swimmers without losing the ability to draw generalized conclusions. The asymmetry requirements of the gold cap for a high velocity of the swimmers are the subject of ongoing theoretical studies.

\section*{Associated Content}\label{Associated Content}
\textbf{Supporting Information} \\
The following files are available free of charge.  \\
Table with parameters of the coating processes. Additional information on gel trapping technique. Calculation of Au cap size of embedded particles. Additional SEM images of Janus particles with different detectors. Additional EDX analysis images. AFM images of particle surface. Mean squared displacement (MSD) curves of all particles at different laser intensities. Corresponding trajectories of the Janus particles in the xy-plane. Diffusion coefficients of all Janus samples. Additional curves of $v_{\text{th}}$ in dependence of the chromium thickness and the sputter rate. (PDF)

\section*{Author Information}\label{Author Info}
\textbf{Author Information} \\
* E-Mail: klitzing@smi.tu-darmstadt.de \\
\textbf{Author Contributions}
All authors have given approval to the final version of the manuscript.  \\
\textbf{Notes} \\
The authors declare no competing financial interest.

\section*{Acknowledgement}\label{Acknowledgement}
The authors greatly thank Mohan Li (Materials Science department, GSI Darmstadt, Germany) for additional assistance with SEM measurements. The authors would like to express their gratitude to the Materials Department at GSI Helmholtzzentrum für Schwerionenforschung for providing the Sputter Coater S150B, to the Materials Department Technical University of Darmstadt (Germany) for providing the Q300TD Sample Preparation System. Many thanks also to Dominik Richter (group of Prof.\,Dr.\,Annette Andrieu-Brunsen, Chemistry Department Technical University of Darmstadt (Germany)) for his help with the thermal evaporation of some Janus particles in the thermal evaporator CREAMET 300 V2. Sincere thanks also to the Target Laboratory at the GSI Helmholtzzentrum für Schwerionenforschung (Darmstadt, Germany) for coating some particles in their BOC Edwards Auto 500 Sputter coater.


\bibliography{references.bib}
\newpage

\end{document}


\beginsupplement
\maketitle

\newpage

\begin{landscape}
\noindent Table\,\ref{SI-table:Parameters} summarizes important parameters (such as the type of the coating device, the coating rates, additional parameters, and the distance between substrate and coating source) of the fabrication process of the Janus particles.

\begin{table}[H]
\caption{Parameters of the coating processes. The first three lines summarize the sputter coaters with ascending gold (Au) sputter rates, while the last line resumes the values for the thermal evaporator.}
\centering
\begin{tabular}{llcccc}
\hline
Sample & Type           & Coating rate /  & Time /  & Parameter & Distance to  \\
& & nm/min & s & & source / mm \\
\hline
Sc5 & Q300TD Sample Prep. System & 25 (Au) & 120 \textsuperscript{a} & 30 mA (Au) & 45 \\
& & 10 (Cr) & 30 \textsuperscript{b} & 60 mA (Cr) &  \\
 & &  &  &  &    \\
Sc1, Em1, Em2, Em3 & Edwards Sputter Coater S150B  & 60 (Au) & 50 \textsuperscript{a} & 30 mA & 50  \\
&&&&&\\
Sc2, Sc3, Sc4& BOC Edwards Auto 500  & 250 (Au) & 12 \textsuperscript{a} & 150 W & 85   \\
&& 30 (Cr) & 10 \textsuperscript{b} & 150 W & 85\\
 & &  &  &  &   \\
Th1 & CREAMET 300 V2  & 30 (Au) & 100 \textsuperscript{a} & - & 270 - 320 \\
&	& 1.2 (Cr)& 250 \textsuperscript{b} & - & 270 - 320 \\
\hline
\multicolumn{5}{l}{\footnotesize{\textsuperscript{a} for a layer of 50 nm Au \textsuperscript{b} for a layer of 5 nm Cr}}
\end{tabular}

\label{SI-table:Parameters}
\end{table}
\end{landscape}

\subsection{1. Gel trapping technique}
Table\,\ref{SI-table:Phytagel} gives an overview of the amount of Phytagel and MgSO$_4$, which was used to create the water subphase for the gel trapping technique. When MgSO$_4$ was added, the appropriate amount of water with MgSO$_4$ inside was added.

\begin{table}[H]
\caption{Amount of Phytagel in the water phase and corresponding concentration of MGSO$_4$.}
\centering
\begin{tabular}{lcc}
\hline
 & m(Phytagel) ; V(H$_2$O) & c(MgSO$_4$)\\
\hline
Em1 & 0.1 g ; 10 ml & - \\
Em2 & 0.01 g ; 10 ml & 4 mM  \\
\hline
\end{tabular}
\label{SI-table:Phytagel}
\end{table}

\noindent 
Since the embedding route, which is described in the experimental section of this paper results only in slightly embedded particles (in the elastomer matrix), it was modified (for Em3).
The liquid silicone elastomer Sylgard184 was poured directly on top of a microscope slide that contains a monolayer of PS particles. After the appropriate curing time of 48 h at room temperature, the silicone elastomer was gently pulled off together with the embedded particles. The Em3 particles were embedded deeper than the Em1 and Em2 particles as indicated by scanning electron microscopy (see Figure\,\ref{fig:SEM}\,(f)). The gold (Au) coating was achieved by subsequent sputter coating of the exposed parts of the particles.
In general, the degree of embedding of the PS particles in the silicone elastomer determines the resulting Au covering. Deeper embedded particles will end up with a smaller Au cap and vice versa.\newline
There are mainly two ways described in the literature to remove the Janus particles from the silicone elastomer matrix: using a sharp blade\cite{Paunov2004} or using adhesive tape and subsequent washing with acetone\cite{Paunov2004,Ye2010}. Both methods are not suitable for the Janus particles in this study. If a sharp blade is used, the coating of some particles could be damaged\cite{Paunov2004}, which we tried to avoid. However, Acetone could also not be used to release the coated Janus particles from the matrix, as this could damage the polystyrene side, e.g. through swelling of the PS particle in acetone\cite{Kumbhar2009}. Other solvents such as ethanol or isopropanol dissolve the adhesive layer only very poorly and incompletely so that sticky residues on the Au side could not be ruled out. In addition, these solvents could hardly remove the Janus particles from the adhesive tape. Instead, a pipette was used to very carefully aspirate the Janus particles under a MilliQ water/isopropanol mixture (add a few drops of isopropanol to 2 ml of water in a Petri dish). This eliminated potential adhesive residues at the Au cap. Excessive pressure during the aspiration process sometimes resulted in larger pieces of still-embedded Janus particles detaching from the matrix (see Figure\,\ref{SI-fig:SEM_SI_Emb}). However, reducing the pressure then resulted in enough individual particles being successfully separated from the matrix material. We then waited ($>$ 24 h) for the small amount of isopropanol to evaporate and further diluted the final dispersion of Janus particles in water with a small amount of MilliQ water, if necessary, until the particle concentration was suitable for our measurements.\\
Since the embedded Em1 particles and the Sc1 particles obtained by sputter coating (both with the same cap size of approx. 74\,\%) display the same thermophoretic velocity (see Figure\,\ref{SI-fig:vth_SI}), potential damages of the Au cap after that removal procedure at least have no discernible influence on the self-propulsion behavior.

\subsection{2. Calculation of Au cap size of embedded particles}
For the embedded particles, the model described in result section of this paper cannot be used to calculate the actual Au cap size since the particles do not shield a certain substrate area. But for these particles, only the part looking out of the embedding matrix can be coated at maximum in any case. For analyses, the internal software features of the Gemini500 FESEM were used, to determine both the diameter of the exposed particle parts and the remaining hole diameter after removal of the particles.\\
The Em3 particles are more than half embedded so that the Au cap size could be determined directly from the exposed particle diameter. For the less than half-embedded particles, one has to analyze the holes in the matrix after particle removal. Since the silicone elastomer matrix is not completely rigid but can be easily deformed, the material slightly collapses after the particles have been removed. It is therefore not possible to infer the Au cap directly from the hole diameters. However, we determined a correction factor based on the hole diameters of the Em3 sample (with an embedding degree of more than 50\,\%), from which one also knows the exposed particle diameters. The hole diameter and the exposed particle diameter at the embedding line must be equal in the situation of a still embedded particle. From the deviation, the factor by which the material collapses is determined. Assuming that this factor does not change too much for the other two embedded samples (which it really shouldn't to any great extent, since the same particles and embedding material were used), one can apply this factor to the rest of the samples and thus back-calculate to the Au cap size. Table\,\ref{SI-table:Embedding} summarizes the corresponding results.

\begin{table}[H]
\caption{Values and assumptions for the calculation of the Au cap size of the embedded particles, obtained from SEM measurements.}
\centering
\begin{tabular}{lccc}
\hline
 & Exposed particle & Empty hole  & Au cap size \\
 & diameter d$_\text{p}$ / \textmu m & diameter d$_\text{h}$ / \textmu m & \\
\hline
Em3 & 2.29 $\pm$ 0.05 & 2.14$^e$ $\pm$ 0.06  & (35.9 $\pm$ 4.0) \%$^d$ \\
&&&\\
\multirow{2}{*}{Em2} & - & (2.15 $\pm$ 0.08) & \multirow{2}{*}{(63.9 $\pm$ 3.1) \%}  \\
 & - & (2.30 $\pm$ 0.12)$^f$ & \\
 &&&\\
\multirow{2}{*}{Em1} & - & (1.16 $\pm$ 0.13) & \multirow{2}{*}{(73.3 $\pm$ 3.0) \%$^h$}  \\
 & - & (1.24 $\pm$ 0.15)$^g$ &   \\
\hline
\multicolumn{4}{l}{\footnotesize{\textsuperscript{d} Au cap size = $\frac{\text{spherical segment covered with gold}}{\text{whole surface area}}$ = $\frac{2 \pi \text{r} \cdot (\text{r}-\sqrt{\text{r}^2-(d_p/2)^2})}{4 \pi \text{r}^2}$ with particle radius r}}\\
\multicolumn{4}{l}{\footnotesize{\textsuperscript{e} Correction factor since the holes shrink after particle removal due to the softness of the silicone }}\\
\multicolumn{4}{l}{\footnotesize{\textsuperscript{  } elastomer material. The hole diameter before particle removal should be equal to the exposed}} \\
\multicolumn{4}{l}{\footnotesize{\textsuperscript{  } particle diameter, so the correction factor is: d$_\text{p}$/d$_\text{h}$ = 1.06923}} \\
\multicolumn{4}{l}{\footnotesize{\textsuperscript{f} exact non-rounded values with correction factor: d$_\text{h}$ = 1.06923 $\cdot$ 2.14683\,$\mu$m = 2.29546 $\mu$m}}\\
\multicolumn{4}{l}{\footnotesize{\textsuperscript{g} exact non-rounded values with correction factor: d$_\text{h}$ = 1.06923 $\cdot$ 1.15690\,$\mu$m = 1.23699 $\mu$m}}\\
\multicolumn{4}{l}{\footnotesize{\textsuperscript{h} Degree of embedding so small that even if particle is obtained by pure sputter coating, the}}\\
\multicolumn{4}{l}{\footnotesize{\textsuperscript{  } corresponding embedded part would never be coated; same value as for Sc1 since the particles are}}\\
\multicolumn{4}{l}{\footnotesize{\textsuperscript{  } only slightly embedded and the same sputter coater was used}}\\
\end{tabular}
\label{SI-table:Embedding}
\end{table}

\newpage
\subsection{3. Scanning Electron Microscopy and Energy Dispersive X-Ray Spectroscopy}
\begin{figure}[H]
\centering
\includegraphics[scale=1.5]{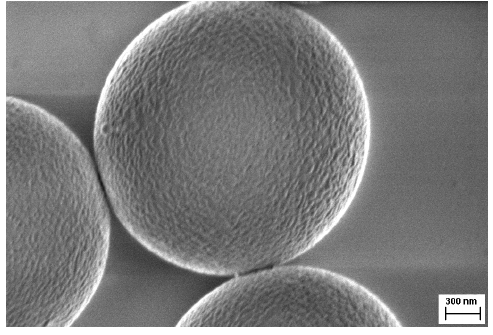}
\caption{SEM image of pure PS particles, which exhibit also a certain surface roughness.}
\label{SI-fig:Pure_PS}
\end{figure}

Figure\,\ref{SI-fig:Pure_PS} shows a SEM image of uncoated PS particles. They also exhibit a certain surface roughness.

\begin{figure}[H]
\centering
\includegraphics[scale=0.5]{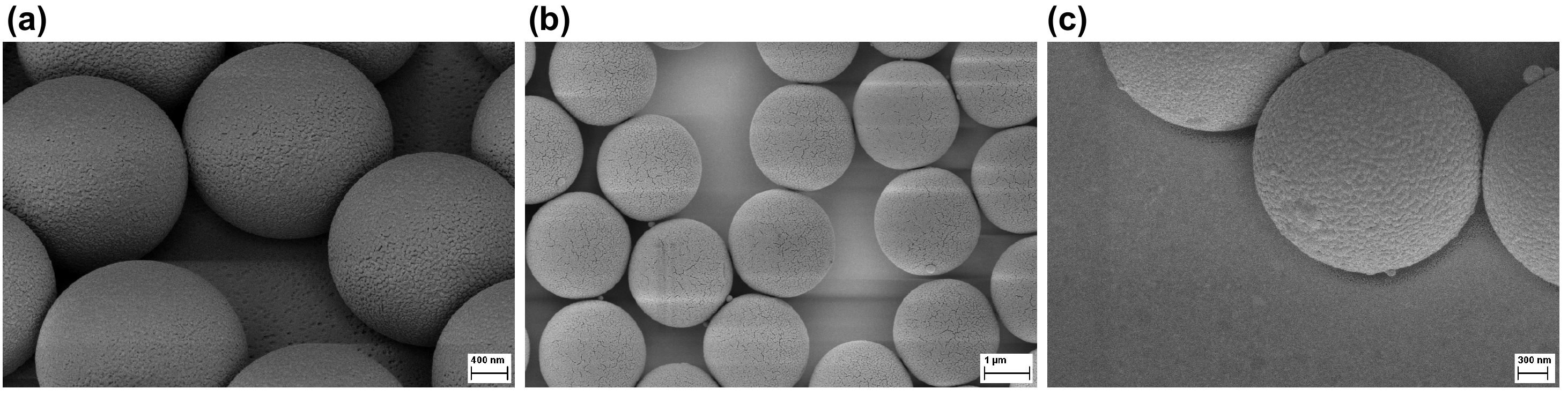}
\caption{Selected SEM images of Janus particles (a) Sc3, (b) Sc4, and (c) Sc5 obtained by sputter coating.}
\label{SI-fig:SEM_SI}
\end{figure}

\begin{figure}[H]
\centering
\includegraphics[scale=0.85]{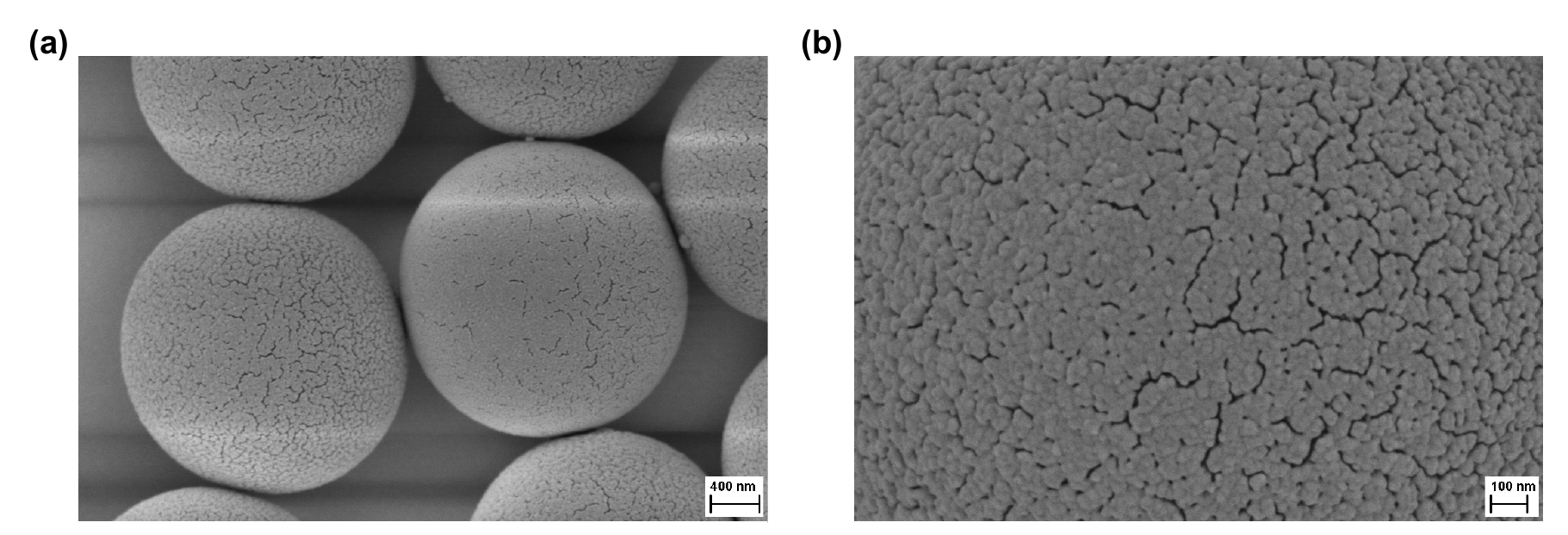}
\caption{Surface structure of Janus particles obtained by sputter coating (a) and enlarged view of single particle surface (b).}
\label{SI-fig:Structure}
\end{figure}

\begin{figure}[H]
\centering
\includegraphics[scale=1.4]{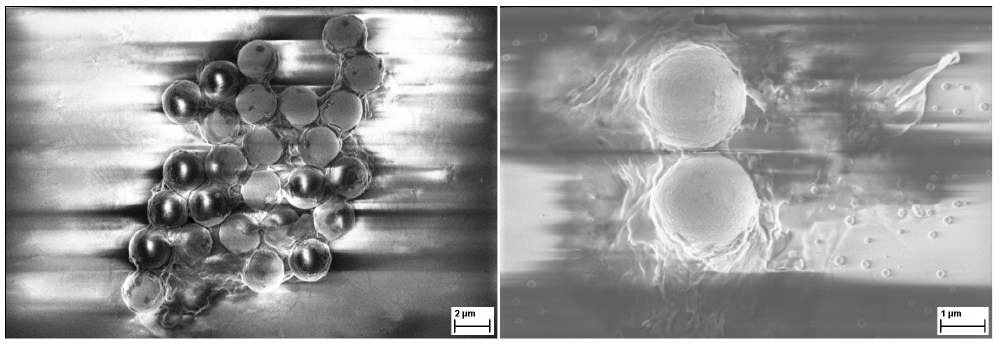}
\caption{SEM images of still embedded Janus particle clusters detached from the matrix.}
\label{SI-fig:SEM_SI_Emb}
\end{figure}

\noindent Figure\,\ref{SI-fig:SEM_SI} summarizes additional selected SEM images of Janus particles obtained by sputter coating, while Figure\,\ref{SI-fig:Structure} displays the corresponding surface structure of a Au cap with Au clusters and trenches in between. Figure\,\ref{SI-fig:SEM_SI_Emb} shows two SEM images where entire clusters of still-embedded Janus particles have detached from the silicone elastomer matrix due to excessive pressure during the detachment process.

\begin{figure}[H]
\centering
\includegraphics[scale=0.4]{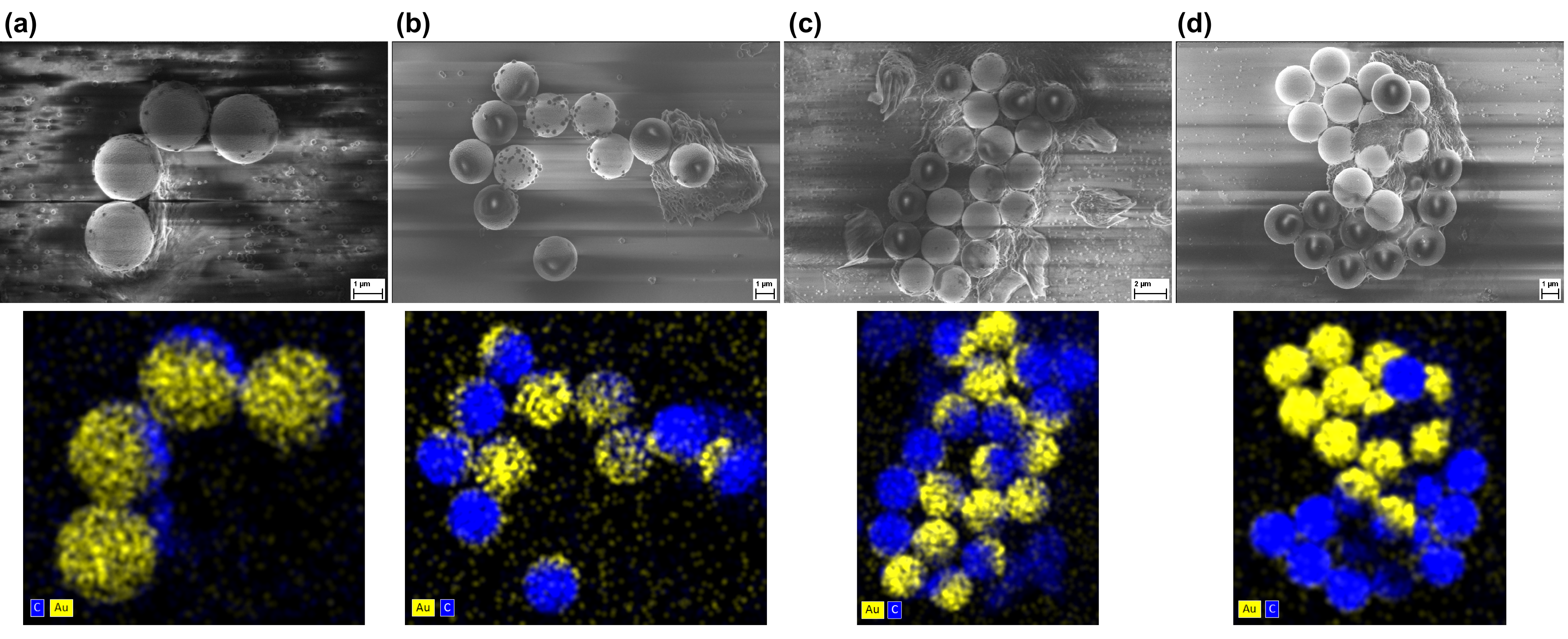}
\caption{Selected SEM images with corresponding results of the EDX analysis of particles obtained by sputter coating. (a) Sc2, (b) Sc3, (c) Sc4, and (d) Sc5, where the yellow part corresponds to the Au cap and the blue part to the PS side.}
\label{SI-fig:EDX_SI_1}
\end{figure}

\begin{figure}[H]
\centering
\includegraphics[scale=0.4]{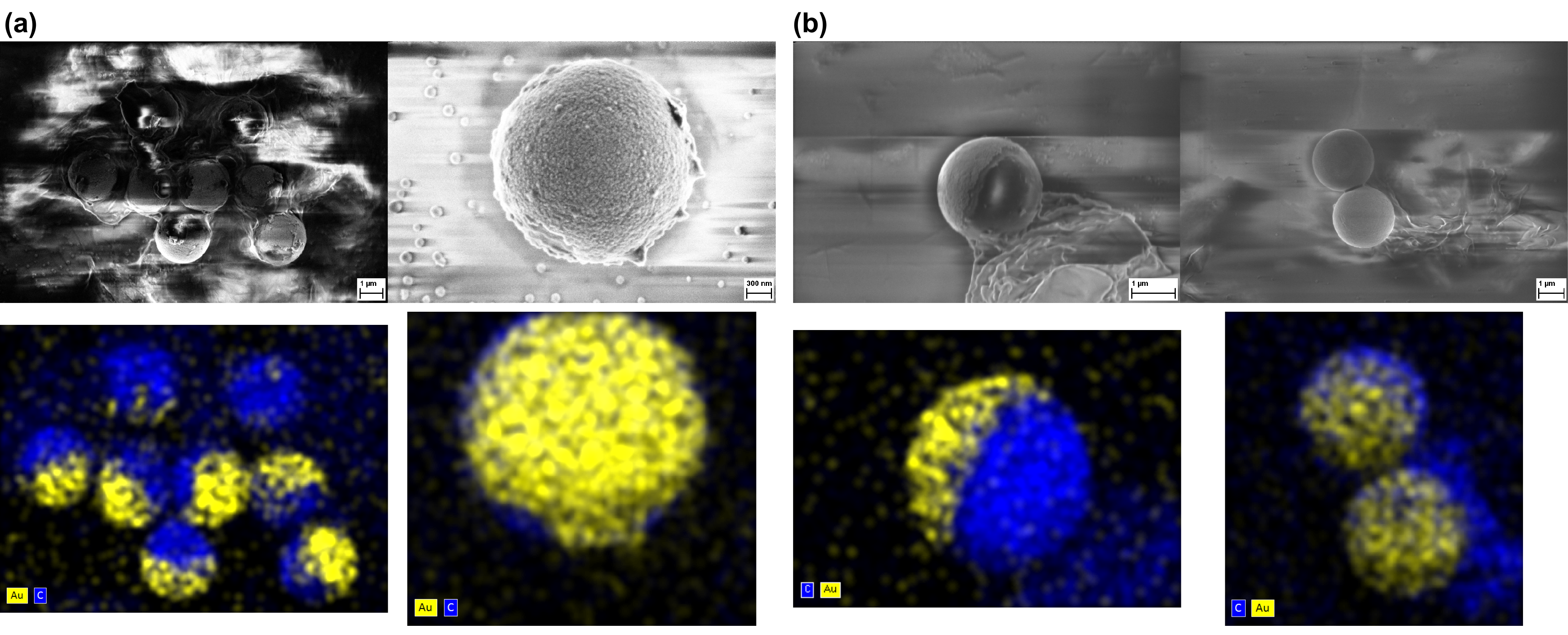}
\caption{Selected SEM images with corresponding results of the EDX analysis of embedded particles (a) Em1 and (b) Em2, where the yellow part corresponds to the Au cap and the blue part to the PS side.}
\label{SI-fig:EDX_SI_2}
\end{figure}

\noindent In Figures\,\ref{SI-fig:EDX_SI_1} and \ref{SI-fig:EDX_SI_2}, selected SEM images with the corresponding EDX analyses can be seen. The blue area maps carbon (i.e., the PS side) and the yellow area maps Au (i.e., the Au cap).

\begin{figure}[H]
\centering
\includegraphics[scale=0.6]{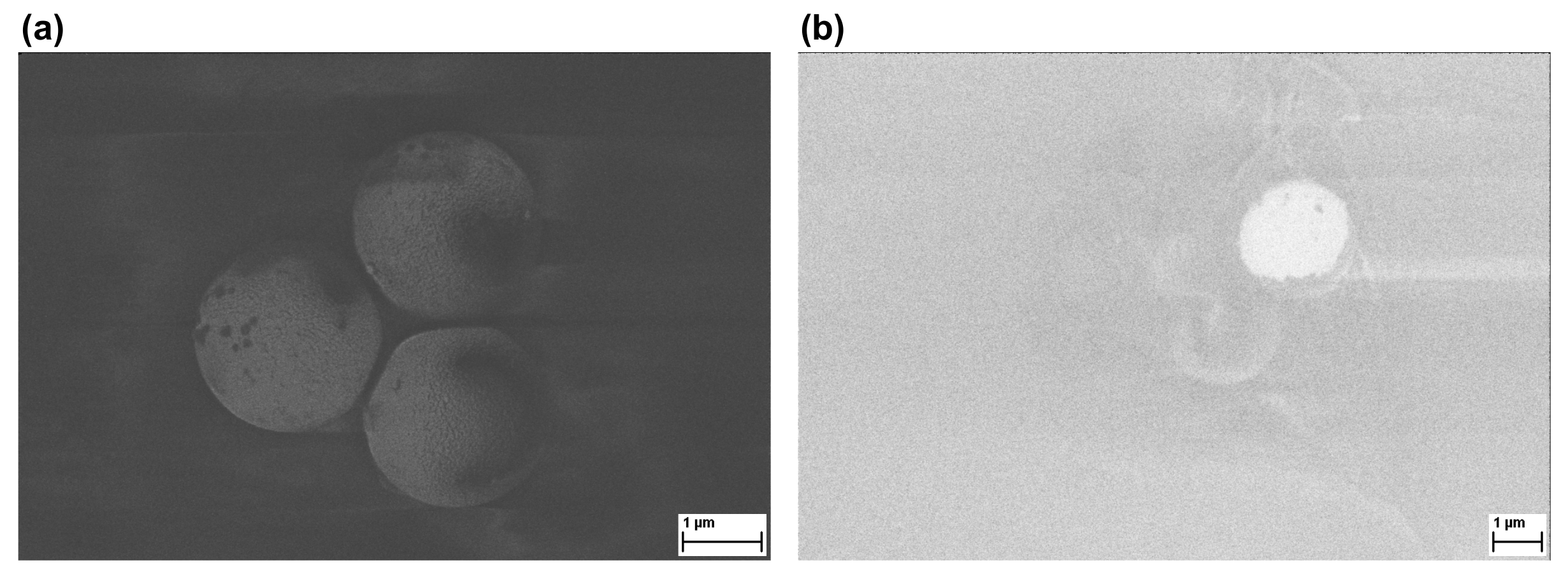}
\caption{Corresponding SEM images of Figure\,\ref{fig:EDX} of the Result part of this paper with different detector (ESB), where the lighter parts indicate the Au cap. (a) Sc1 particles and (b) Em3 particles.}
\label{SI-fig:SEM_SI_Detector}
\end{figure}

\begin{figure}[H]
\centering
\includegraphics[scale=0.5]{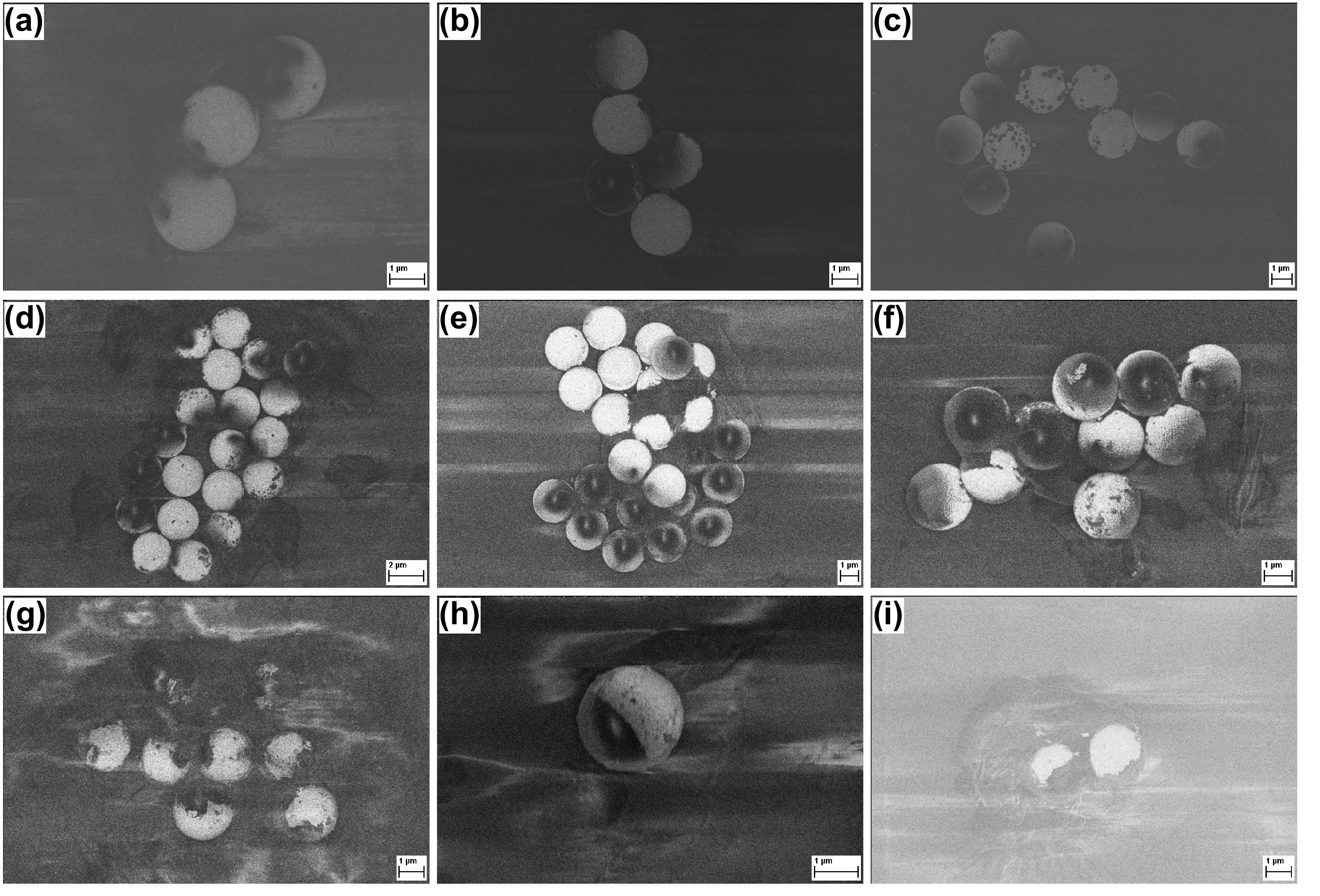}
\caption{SEM images of all analyzed particles with different detector (ESB), where the lighter parts indicate the Au cap. (a) Sc1 particles, (b) Sc2, (c) Sc3, (d) Sc4, (e) Sc5, (f) Th1, (g) Em1, (h) Em2, and (i) Em3.}
\label{SI-fig:SEM_SI_Detector_2}
\end{figure}

\noindent Additional SEM images were also recorded using an ESB (Energy selective backscattered) detector. This enhances the contrast between uncoated PS-side and Au cap. The corresponding results are shown in Figure\,\ref{SI-fig:SEM_SI_Detector} and Figure\,\ref{SI-fig:SEM_SI_Detector_2}.

\subsection{4. Atomic Force Microscopy}
\begin{figure}[H]
\centering
\includegraphics[scale=0.6]{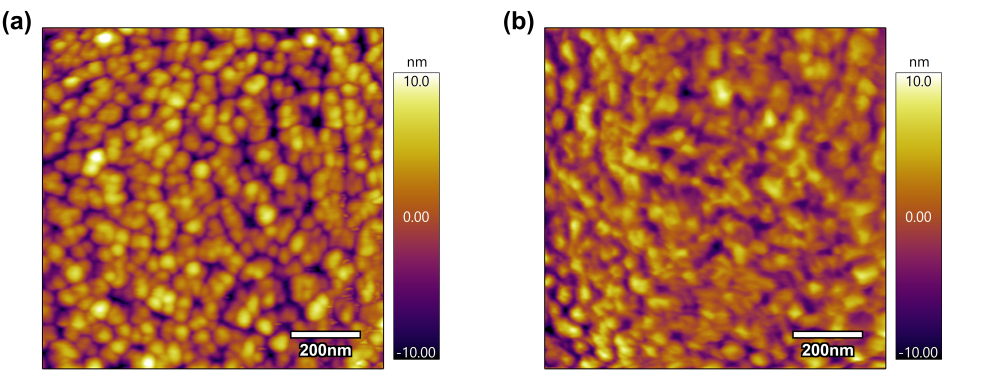}
\caption{AFM image of surface structure of Au cap obtained with (a) sputter coating and (b) thermal evaporation.}
\label{SI-fig:surface_structure}
\end{figure}

AFM measurements in tapping mode were also used to determine the Au cap roughness of the prepared Janus particles. Figure\,\ref{SI-fig:surface_structure} shows the surface topography of a single Janus particle in ambient conditions prepared via sputter coating (a) and thermal evaporation (b). The surface topography of all sputter-coated Au caps appeared similar irrespective of the used sputter coating machine or sputter rate. The surface features of the particles obtained by thermal evaporation revealed the presence of very similar structures.

\subsection{5. Thermophoretic Self-Propulsion}
\begin{figure}[H]
\centering
\includegraphics[scale=0.8]{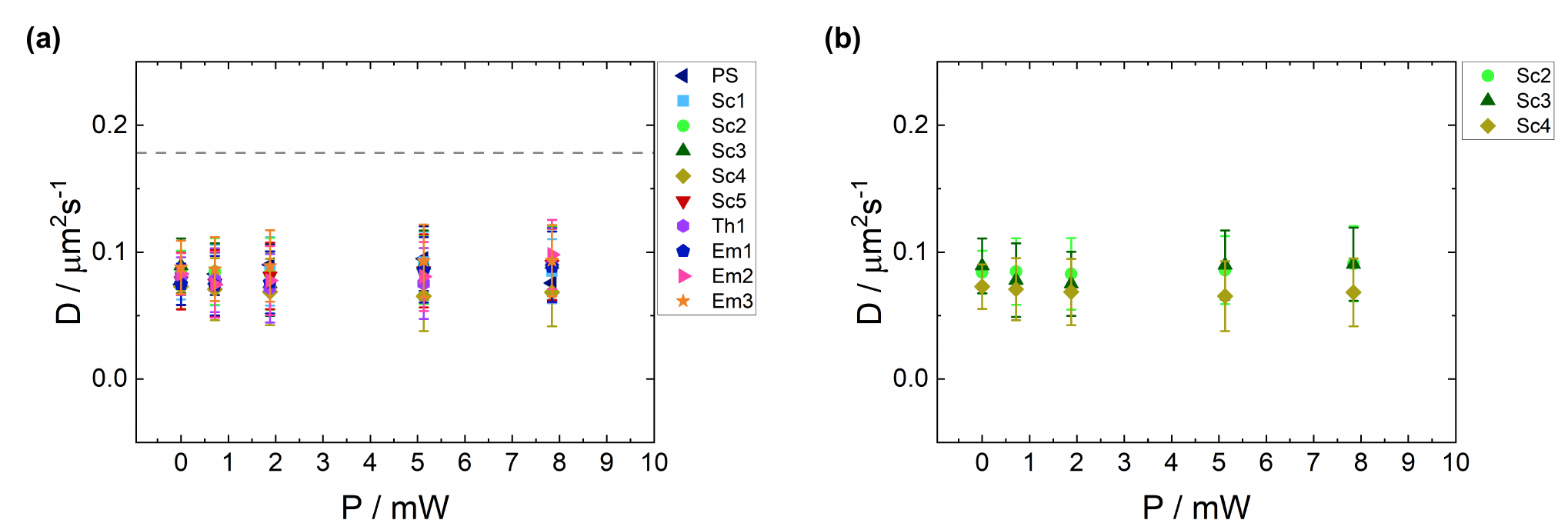}
\caption{(a) Diffusion coefficient of all analyzed particles at different laser powers. Diffusion coefficient for Janus particles with increasing chromium layer thickness from 0\,mW (Sc2), 5\,mW (Sc3) to 50\,mW (Sc4). The dotted line in (a) corresponds to D$_{\text{bulk}}$.}
\label{SI-fig:Diffusion_coefficient}
\end{figure}

\begin{figure}[H]
\centering
\includegraphics[scale=1.5]{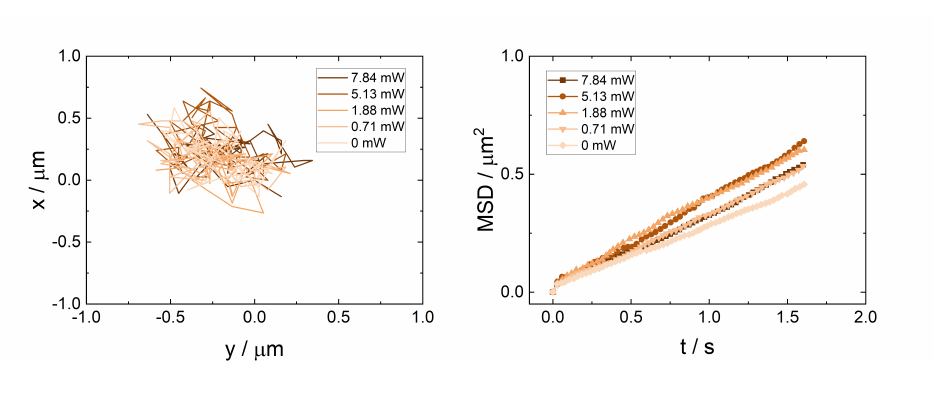}
\caption{Trajectories and MSD curves of the pure PS particles.}
\label{SI-fig:MSD_traj3}
\end{figure}

\begin{figure}[H]
\centering
\includegraphics[scale=1.2]{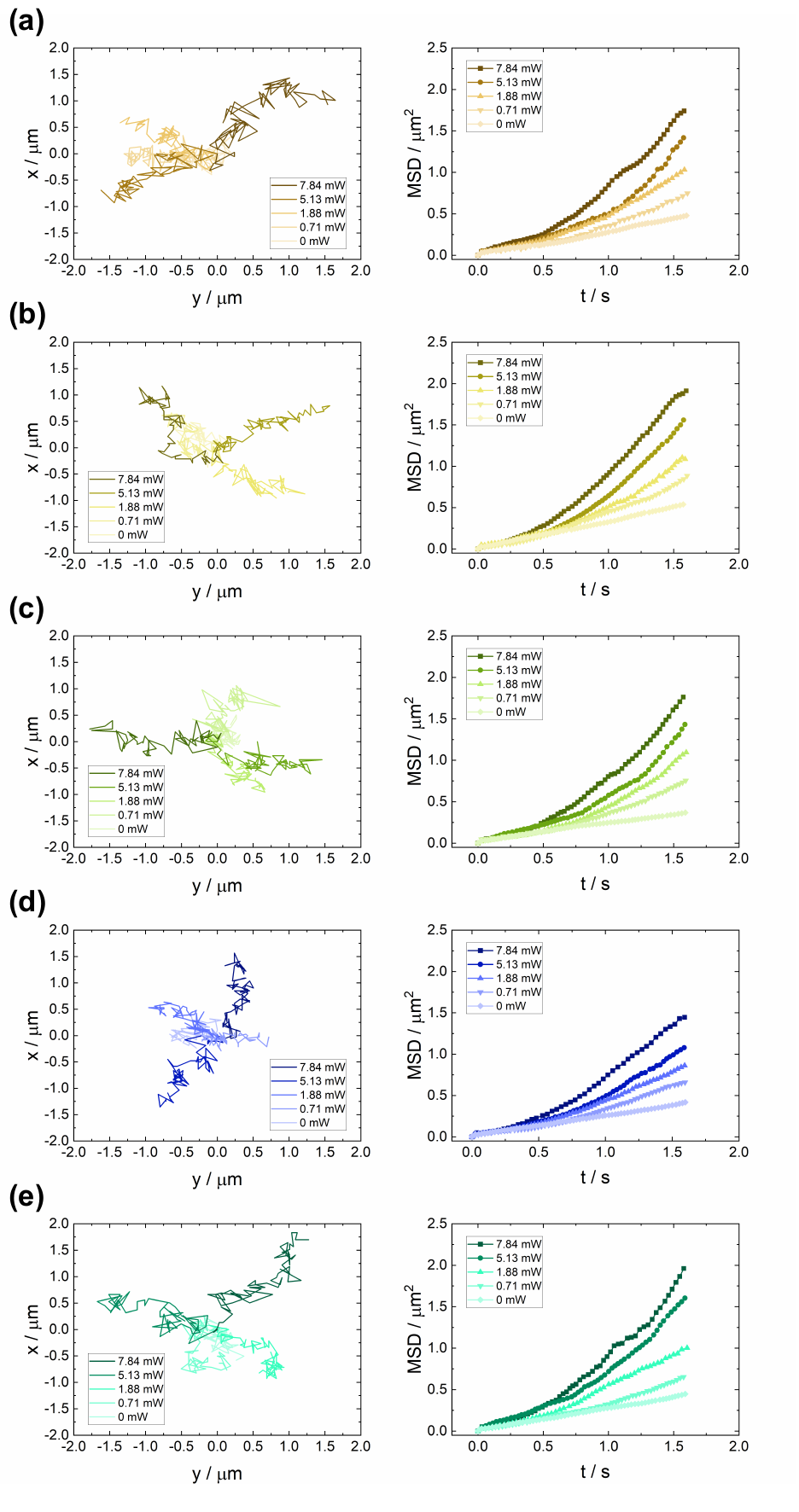}
\caption{Trajectories and MSD curves of the Au-PS Janus particles obtained by sputter coating: (a) Sc1, (b) Sc2, (c) Sc3, (d) Sc4, and (e) Sc5.}
\label{SI-fig:MSD_traj}
\end{figure}

\begin{figure}[H]
\centering
\includegraphics[scale=1.2]{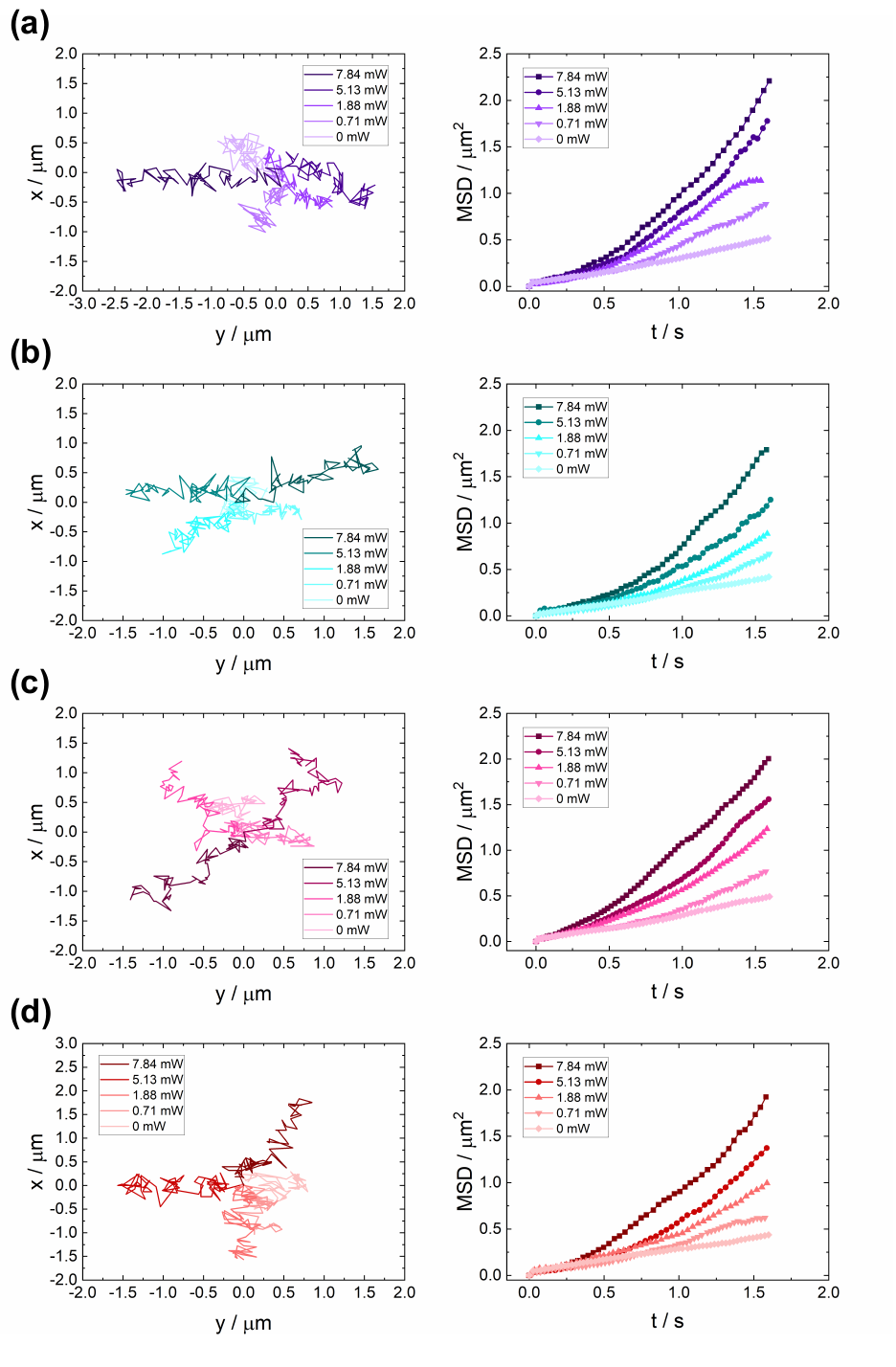}
\caption{Trajectories and MSD curves of the Au-PS Janus particles obtained by (a) thermal evaporation (Th1) and embedding and sputter coating: (b) Em1, (c) Em2, and (d) Em3.}
\label{SI-fig:MSD_traj2}
\end{figure}

\begin{figure}[H]
\centering
\includegraphics[scale=0.55]{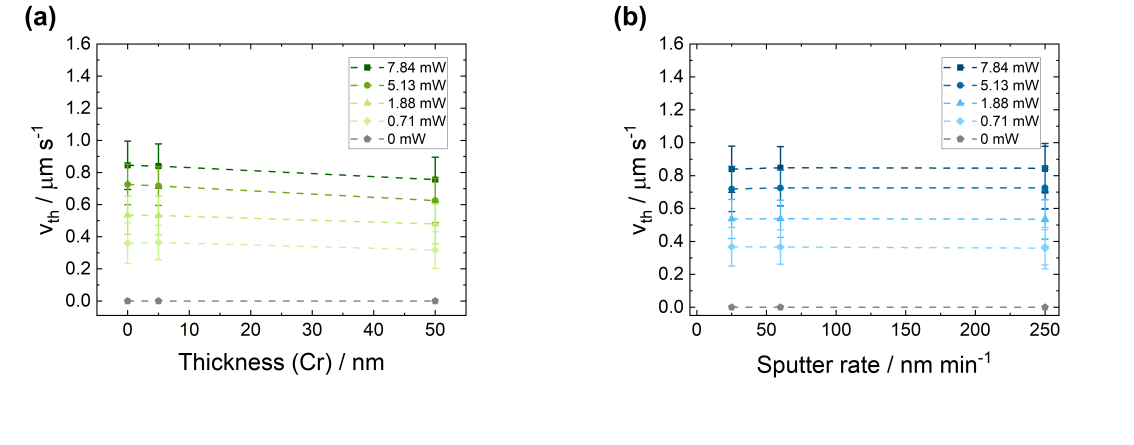}
\caption{Thermophoretic velocity in dependence of (a) the chromium thickness and (b) the sputter rate.}
\label{SI-fig:vth_SI}
\end{figure}

The diffusion coefficients of all Janus particles are summarized in Figure\,\ref{SI-fig:Diffusion_coefficient}\,(a). Figure\,\ref{SI-fig:Diffusion_coefficient}\,(b) shows only the diffusion coefficients of Sc2 (50\,nm Au), Sc3 (50\,nm Au + 5\,nm Cr), and Sc4 (50\,nm Au + 50\,nm Cr). Figure\,\ref{SI-fig:MSD_traj3} summarized the trajectories and the MSD curves of the pure PS particles. Figures\,\ref{SI-fig:MSD_traj} and \ref{SI-fig:MSD_traj2} display the trajectories and corresponding MSD curves of the analyzed Janus particles, while Figure\,\ref{SI-fig:vth_SI} shows the dependence of $v_{\text{th}}$ on the chromium layer thickness (a) and sputter rate (b).

\bibliography{references.bib}